\documentclass[smallextended]{svjour3}
\usepackage[bookmarks=false]{hyperref}
\usepackage{color}
\usepackage{url}
\usepackage{wrapfig}
\usepackage{algorithm}
\usepackage{algpseudocode}
\usepackage{graphicx}
\usepackage{enumitem}
\usepackage{epsfig}
\usepackage[dvipsnames]{xcolor}
\usepackage{latexsym, amsmath, amsfonts, amssymb, bbm}
\usepackage{balance}
\usepackage{natbib}
\usepackage{todonotes}
\usepackage{booktabs}
\usepackage[normalem]{ulem}
\useunder{\uline}{\ul}{}
\usepackage{longtable}
\usepackage{multirow}

\usepackage{caption}
\usepackage[most]{tcolorbox}

\newcommand{\sectopic}[1]{\vspace{0em}\par\noindent{\textit{\bfseries #1}}}

\usepackage{etoolbox}
\newcommand*{\affaddr}[1]{#1} 
\newcommand*{\affmark}[1][*]{\textsuperscript{#1}}

\begin{document}

\title{Managing Human-Centric Software Defects: Insights from GitHub and Practitioners' Perspectives
}


\author{
		Vedant~Chauhan\affmark[1] \and
		Chetan~Arora\affmark[1]  \and
        Hourieh~Khalajzadeh\affmark[2] \and
        John~Grundy\affmark[1] 
}

\institute{Vedant Chauhan \and Chetan Arora \and Hourieh Khalajzadeh \and John Grundy \at              \email{vedant.chauhan@monash.edu, chetan.arora@monash.edu, \\ hkhalajzadeh@deakin.edu.au, john.grundy@monash.edu}           \\
              {Tel.: +61 3 9905 8854}          \\ \\
            \affaddr{\affmark[1]Faculty of Information Technology, Monash University, Melbourne, Australia}\\
\affaddr{\affmark[2]School of Information Technology, Deakin University, Geelong, Australia}
}

\date{Received: date / Accepted: date}

\maketitle

\thispagestyle{plain}

\pagestyle{plain}

\begin{abstract}
\textit{Context}: Human-centric defects (HCDs) are nuanced and subjective defects that often occur due to end-user perceptions or differences, such as their genders, ages, cultures, languages, disabilities, socioeconomic status, and educational backgrounds. Development teams have a limited understanding of these issues, which leads to the neglect of these defects. Defect reporting tools do not adequately handle the capture and fixing of HCDs.

\textit{Objective}: This research aims to understand the current defect reporting process and tools for managing defects. Our study aims to capture process flaws and create a preliminary defect categorisation and practices of a defect-reporting tool that can improve the reporting and fixing of HCDs in software engineering.

\textit{Method}: We first manually classified 1,100 open-source issues from the GitHub defect reporting tool to identify human-centric defects and to understand the categories of such reported defects. We then interviewed software engineering practitioners to elicit feedback on our findings from the GitHub defects analysis and gauge their knowledge and experience of the defect-reporting process and tools for managing human-centric defects.

\textit{Results}: We identified 176 HCDs from 1,100 open-source issues across six domains: IT-Healthcare, IT-Web, IT-Spatial, IT-Manufacturing, IT-Finance, and IT-Gaming. Additionally, we interviewed 15 software practitioners to identify shortcomings in the current defect reporting process and determine practices for addressing these weaknesses.

\textit{Conclusion}: HCDs present in open-source repositories are fairly technical, and due to the lack of awareness and improper defect reports, they present a major challenge to software practitioners. However, the management of HCDs can be enhanced by implementing the practices for an ideal defect reporting tool developed as part of this study.
\end{abstract}


\section{Introduction}
\label{sec:intro}

Software defects are a significant problem affecting the overall functionality and dependability of software systems, posing a challenge to developers, quality assurance teams, and end-users. User preferences and differences when interacting with software systems are a critical influence in diagnosing and mitigating software defects \citep{strate2013literature,yusop2016reporting}. These defects can appear as a result of varied software usage from different users' perspectives. 
\citep{huynh2021improving}. 

\begin{figure*}
\centering
  \includegraphics[width=\textwidth]{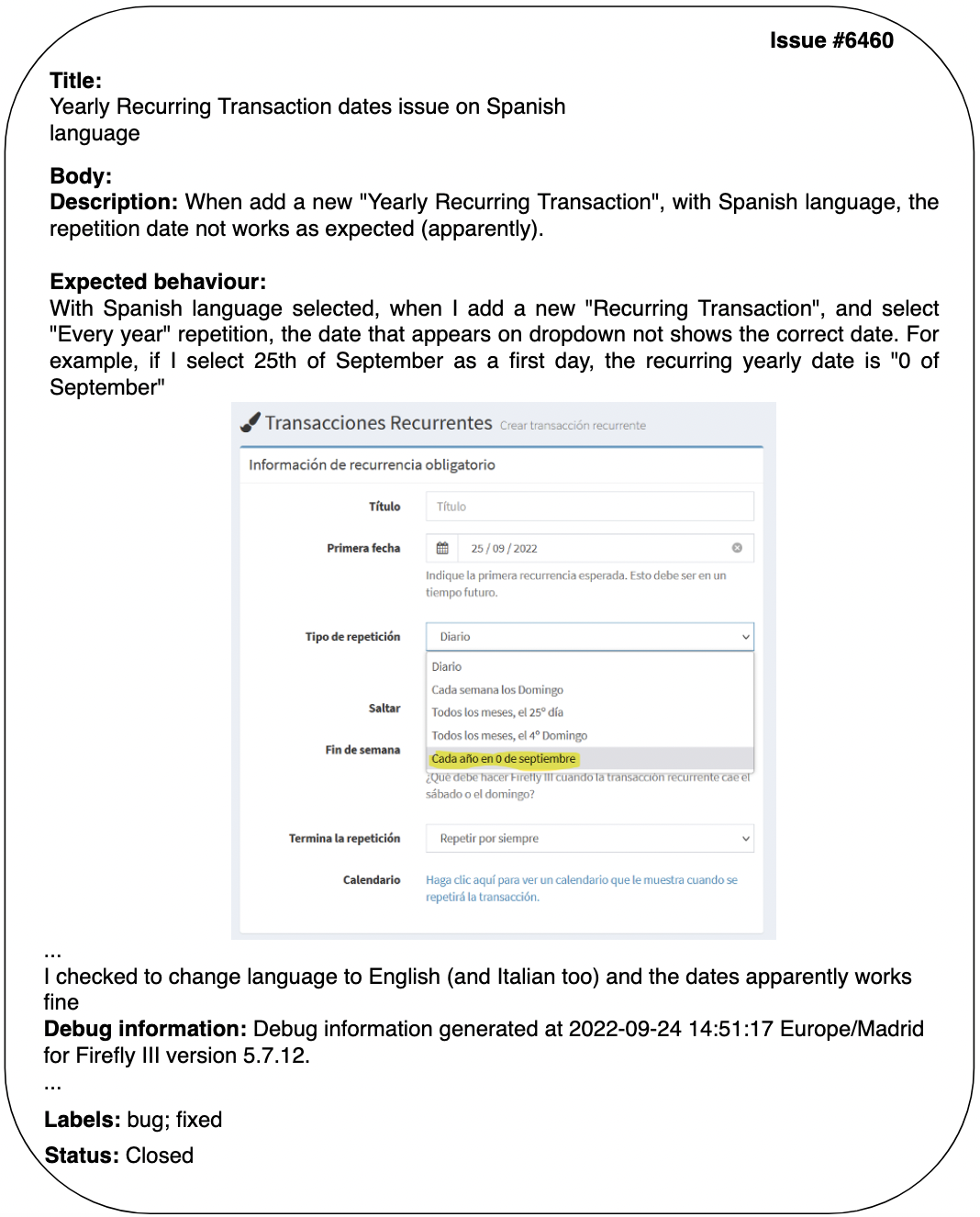}
\caption{An example HCD reported in firefly-iii GitHub Repository}
\label{fig:hcdeggithub}       
\end{figure*}

Our research focuses on what we term ``human-centric defects" (HCDs). We define HCDs as nuanced and often subjective defects that occur when different end-users interact with the application. During our research, we found that HCDs in the SE domain occur due to end-user perception or differences, such as their technical knowledge, privacy, educational background, age, gender, location, translation, language, and application causing bugs specific to them. These defects can be considered outliers or edge conditions missed by the development teams while developing or testing their products. \citep{chauhan2024software}. Fig.~\ref{fig:hcdeggithub} showcases a real-world example of HCD reported in a finance open-source project called `firefly-iii'\footnote{\label{note1}https://github.com/firefly-iii/firefly-iii/issues/6460}. The issue highlights an application that is not allowing a user to add a new ``Yearly Recurring Transaction", in Spanish language. However, the same feature is working with English or Italian language. The user created an issue in Firefly-iii's GitHub repository and provided additional debugging context for the developer. The issue report further consists of a basic description, expected behaviour, debug information, and screenshots. The report also has some missing defect report fields such as system, observed behaviour, and additional context regarding the issue, which can make triaging the issue difficult. As noted above, this is a subjective issue or for a specific user group and, hence, a non-issue for user groups that do not require the Spanish language for transaction dates in their activities. Hence, it can be categorised as an HCD in the usability defect category due to language issues in UI. 

 Defects are commonly categorised as either functional or non-functional. Functional defects disrupt the functioning of the system, namely the functional requirements. Non-functional defects, on the other hand, occur when the quality of the system is compromised, particularly non-functional requirements such as usability, security, performance, and compatibility flaws \citep{matrinnfr, eckhardt2016non}. We define ``normal technical defects" as software flaws that are applicable to all users and usually stem from the absence of certain features, errors made by developers, or ambiguous business requirements from both functional and non-functional sides. HCDs can potentially be classified under both of these defect types. When ordering an Uber, the process of selecting a child seat can be perplexing and demands more effort from the consumer to ensure it is included in their ride. Furthermore, it is not universally accessible in all areas, and the availability of car seats may differ within a specific city. This is not a concern for a user not looking to book a ride with a child, but only for specific user groups, e.g., parents with children who require a child seat. This issue can be categorised as a functional problem in the application, specifically the absence of a feature to reserve a child seat. A non-functional issue occurs, for instance, when a user's colour blindness affects their ability to access the app's features fully~\citep{chauhan2024software}, e.g., inadequate support for navigation on maps for visually impaired users for selecting a location of their ride or pickup.

Identifying HCDs at an earlier stage requires constant user engagement and considering their diverse requirements, diverse development teams, and awareness. A number of studies \citep{fazzini2022characterizing,grundy2020human,hidellaarachchi2021effects,grundy2020humanise,grundy2018vision} establish the challenges of gaining access to diverse users, capture human-centric issues, absence of adequate feedback to developers, lack of adoption strategies, effort, and sustainability. Defect reporting tools exhibit a failure to accurately handle HCDs when a user reports a defect. The defect reports generated by these tools often lack necessary information, are ambiguous, and exhibit a broad range in quality due to various reasons \citep{yusop2016reporting,bettenburg2008makes,yusop2017analysis,chauhan2024software}. The quality of defect reports may be compromised by a deficiency in understanding the bug reporting procedure, tools, and expertise in reporting defects \citep{ko2010power}, thereby reducing the overall quality of the reports. An inconsistency in information can arise when reporting bugs to developers, where there is a discrepancy between what is expected by the developers and what is actually provided to them \citep{bettenburg2008makes}. The information presented may occasionally lack lucidity, be partial, or lack comprehensive details \citep{davies2014s}. In addition, the defect reports lack sufficient recommendations for problem fixes \citep{hornbaek2006kinds}, and do not provide coherent and logical textual details or traced information \citep{strate2013literature}. Furthermore, unclear defect reports create a communication gap between developers and testers \citep{wang2011really}. 

Due to the challenging nature of reporting HCDs and a lack of awareness, we conducted a previous study to comprehend the perception of human-centric defects in the software industry~\citep{chauhan2024software}. Our study determined that HCDs are distinct and software organisations and teams lack sufficient awareness of human-centric factors, resulting in HCDs being disregarded in software development. The results of our investigation suggest that the existing defect reporting tools are inadequate for managing HCDs; this can be remedied by incorporating automation support, an efficient end-user feedback system, and a more comprehensive taxonomy. Our study further helped narrow down the focus of HCDs into five key categories and their relative importance based on practitioners' viewpoints. These five categories, in their sequence of perceived importance (higher to lower), are usability, functionality, compatibility, performance, and security. This study is an extension of our previous research. Specifically, this study explores the classification of HCDs in real reported defects and the management of HCDs by practitioners using existing defect-reporting tools.
To this end, we manually classified 1,100 defects from GitHub\footnote{https://github.com} defect reporting tool into five key HCD categories to investigate whether the perceived importance of these defect categories matches their reported frequency. We also interviewed 15 participants to cross-validate our findings from the classification and to understand their current practices and challenges faced while reporting HCDs using existing defect reporting tools. Our study makes the following key contributions:

\begin{itemize}
    \item formulating a preliminary defect categorisation of HCDs found in different GitHub open-source repositories;
    \item  an analysis of existing defect reporting processes and defect reports for handling HCDs;
    \item  effective integration of the diverse human perspective into defect reporting tools; and 
    \item presenting a set of practices for defect reporting tools to efficiently manage HCDs.
\end{itemize}

\sectopic{Structure.} The rest of the paper is structured as follows. Section~\ref{sec:relatedwork} positions our work against the related work. Section~\ref{sec:rm} provides an insight into our research methodology, i.e., our research questions (RQs), and our data collection process for classification and interviews. Results and discussion from our classification and interviews are provided in Sections~\ref{sec:results} and \ref{sec:discussion}. Section~\ref{sec:threats} presents threats to the validity, and Section~\ref{sec:conclusion} concludes the paper.


\section{Related Work}
\label{sec:relatedwork}

There are multiple studies conducted on software repositories and issue-reporting systems to understand and improve the state of human aspects encountered in such repositories. \citet{strate2013literature} performed a comprehensive analysis of software defect reporting, focusing on five specific areas of defect reporting: triaging, quality, metrics of defect reports, automatic defect fixing, and defect detection. Triage is typically a tedious task, and the speed at which an issue is resolved is influenced by the quality and metrics of defect reports. The focus of automatic bug detection and resolution is on reducing manual labour in the process and exploring breakthroughs in the sector to expedite problem resolution. \citep{ko2011design} explored open-source bug reports to identify the conflicts between user requirements and intended design challenges. Some studies \citep{yusop2017analysis,twidale2005exploring,andreasen2006usability} explored usability defect reports to identify developers’ perspectives on usability interaction and evaluation. Usability defects occur when a user encounters unintended behaviour while using a software application. These bugs are reported by both expert and non-expert users. This is similar to HCDs; however, HCDs occur in a larger subset of defect categories, such as functional and non-functional defects. \citep{dabbish2012social} investigated transparency in collaboration and management in communities of practice in GitHub users while \citep{dam2015mining} mined software repositories to extract social norms. \citep{barcellini2008socio} analysed the Open Source Software's social, design, and thematic temporal dynamics (OSS) to understand how knowledge and artefacts are magnified in the OSS community. Although these findings serve as an initial step, further research exploration on HCDs is necessary.

Analysis of developers’ human characteristics from software repositories and defect-reporting systems has been explored by some studies. \citep{ortu2015jira} investigated the Jira\footnote{https://jira.atlassian.com} issue reporting system for four open-source projects and comprehended the developer communication process in such projects, identifying the emotions and sentiments in the developer’s comments. The positive emotions of developers showcased that the time taken to fix the issue is less, while the negative emotions increased the time to fix the issue. Emotion classification is also explored by \citep{cabrera2020classifying} for StackOverflow\footnote{https://stackoverflow.com} posts and Jira issue comments. They analysed different lexica and label algorithms to classify emotions which showed considerable improvements in emotion classification. Developers’ emotions are also investigated by \citep{murgia2014developers} in the Apache Software Foundation issue reporting system. They highlighted that developers express gratitude, joy, and sadness while discussing issues. An increase in the amount of context in a defect report creates a nuanced understanding of emotions in humans. However, there needs to be more investigation into human aspects to thoroughly construct a concrete taxonomy and automation in this domain.

The issue with the existing defect reporting procedure lies in its inefficiency in capturing HCDs, resulting in prolonged resolution durations or, in some cases, outright neglect of HCDs within the current SE process. It is imperative to investigate the human elements involved in defect monitoring and reporting in the software engineering sector. \citep{khalajzadeh2022diverse} investigated 1691 GitHub issue comments to perform a diverse empirical study to identify eight categories: privacy \& security, location \& language, compatibility, inclusiveness, accessibility, preference, emotional aspects and satisfaction of human-centric issues conferred by developers. \citep{yusop2016reporting}, and \citep{houriehhcdinapps} suggested there is no structured way of reporting human-centric issues in general or in-app reviews. In addition, the structure and typical characteristics, such as description, severity, steps to reproduce, stack traces, and expected behaviour, contribute to resolving issues more efficiently. Awareness regarding the reporting and managing of human aspects is currently limited in many works of literature. \cite{huynh2021improving} provided some insights into preliminary works on human-centric defect reporting. They captured a subset of humans’ underlying conditions via personas such as vision impairment, colour blindness, hearing impairment, dyslexia, dexterity and aphasia. They developed initial prototypes to support human-centric defect reporting based on the personas. While this work is commendable, more such studies need to be conducted to better understand human-centric defect reporting. We require more detailed research on HCD from all different defect reporting areas such as triaging, classification, categorisation reporting, and resolution. The current research addresses some of these aspects. In this study, we are trying to explore the classification, reporting and fixing of HCDs via the defect reporting tools available in the SE domain.


\section{Research Methodology}
\label{sec:rm}

In this section, we present our research questions, the research methodology we used for our classification and interviews, and the demographics of our interview participants.

\subsection{Research Questions}~\label{sec:rq}
We wanted to better understand the management of HCDs in existing software defect reporting processes. To achieve this objective, we formulated the following research questions (RQs):

\sectopic{RQ1. Which HCD categories are the most prominently covered in open-source repositories, such as GitHub?} Based on the results collected in our previous study~\cite {chauhan2024software}, we classify open-source Github issues into five HCD categories, i.e., functional and non-functional (usability, performance, security, and compatibility). Non-functional requirements encompass various categories. This study concentrated on four specific types: usability, performance, security, and compatibility, as referenced in \citep{matrinnfr, eckhardt2016non}.  RQ1 triangulates our findings from the previous study. It verifies whether the claims provided by practitioners on the frequency of reported HCD categories compare with the open-source issue analysis of HCDs conducted as part of RQ1. Additionally, we create a preliminary defect categorisation of HCDs in the defect categories as mentioned above, based on our manual classification effort.

\sectopic{RQ2. Do existing defect-reporting processes and tools accommodate HCDs?} RQ2 identifies the defect reporting process and practices in relation to HCDs. Given that HCDs often directly affect the interaction between users and applications, it is important to investigate how these issues are identified, reported, and prioritised within existing defect reporting tools.

\sectopic{RQ3. Can the reporters’ background impact the handling of HCDs in the defect-reporting process?} This RQ studies the importance of the reporter’s background being integrated into the defect report, and the impact on developers, teams, and reporters in the overall defect reporting process. 

\sectopic{RQ4. What improvements can be applied to effectively assist in the reporting of HCDs in defect reporting tools?} This RQ identifies the practices required in the defect reporting tools to capture HCDs.


We conducted the research in two parts, first, the analysis and classification of the GitHub open-source issues. Second, interviews with SE practitioners to understand their perspectives on managing HCDs in defect reporting tools. Next, we explain our data collection process for classification and interviews.

\subsection{Classification Approach}
\label{sec:classification}

In our previous study \citep{chauhan2024software}, we surveyed 50 and interviewed 10 SE practitioners. During the study, we documented the fields of work or domains of these practitioners, which included IT-Healthcare, IT-Web, IT-Spatial, IT-Manufacturing, IT-Finance, IT-Government, IT-Telecommunications, and IT-Gaming. Given the diversity and relevance of these domains, we conveniently sampled projects from GitHub within these same domains to ensure consistency in our analysis while following the below inclusion criteria:
\begin{itemize}
    \item \textbf{Domain labels:} We compiled a selection of open-source repositories with domain-specific labels. For example, when we searched for repositories labeled ``IT-Healthcare" on GitHub, no projects were available. Given that most projects on GitHub are related to IT, using ``IT-Healthcare" as a label proved to be impractical. Consequently, we opted to use ``healthcare" as the label. This methodology was consistently applied across other domains. Table~\ref{tab:githubrepos} provides further details on the selected projects.
\item \textbf{Number of stars:} Github allows the users to `star' the projects they deem interesting, which allows them to track the progress of these projects. A higher number of stars is a sign of mature projects of interest to users/developers~\citep{schwartz2023five}. We collected open-source repositories with at least 2000 stars.
\item \textbf{Number of issues:} We collected open-source repositories with more than 100 issues reported.
\item \textbf{Language:} We only considered projects and corresponding issues primarily specified in English.
\end{itemize}

\begin{longtable}{lllll}
\caption{List of GitHub projects}
\label{tab:githubrepos}\\
\hline
\textbf{Domain}  & \textbf{\begin{tabular}[c]{@{}l@{}}GitHub \\ Repository\end{tabular}}          & \textbf{\begin{tabular}[c]{@{}l@{}}Number \\ of issues\end{tabular}} & \textbf{\begin{tabular}[c]{@{}l@{}}Number \\ of Stars\end{tabular}} & \textbf{Labels}                                                                                                                                                                                                                                                                                   \\ \hline
\endfirsthead
\endhead
IT-Spatial       & \begin{tabular}[c]{@{}l@{}}tidwall/\\ tile38\end{tabular}                      & 546                                                                  & 8,900                                                               & \begin{tabular}[c]{@{}l@{}}database;location;geo;\\ geospatial;spatial;\\ geofences;index\end{tabular}                                                                                                                                                                                            \\ \hline
IT-Spatial       & \begin{tabular}[c]{@{}l@{}}gboeing/\\ osmnx\end{tabular}                       & 636                                                                  & 4,600                                                               & \begin{tabular}[c]{@{}l@{}}python;mapping;\\ openstreetmap;osm;\\ transportation;geospatial;\\ routing;gis;spatial;\\ urban-planning;transport;\\ networkx;networks;\\ spatial-analysis;overpass-api;\\ spatial-data;geography;\\ street-networks;urban;osmnx\end{tabular}                        \\ \hline
IT-Gaming        & \begin{tabular}[c]{@{}l@{}}pixijs/\\ pixijs\end{tabular}                       & 5,466                                                                & 42,300                                                              & \begin{tabular}[c]{@{}l@{}}javascript;game;webgl;\\ canvas;rendering;glsl;\\ data-visualization;renderer;\\ pixijs;pixi;canvas2d;\\ rendering-engine;\\ rendering-2d-graphics\end{tabular}                                                                                                        \\ \hline
IT-Gaming        & \begin{tabular}[c]{@{}l@{}}dkhamsing/\\ open-source-\\ ios-apps\end{tabular}   & 278                                                                  & 39,800                                                              & \begin{tabular}[c]{@{}l@{}}game;swift;ios;app;list;\\ apple;awesome;react-native;\\ objective-c;tvos;watchos;\\ cocoapods;example;iphone;\\ ipad;apple-tv;apple-watch;\\ flutter;swiftui;apple-vision-pro\end{tabular}                                                                            \\ \hline
IT-Manufacturing & \begin{tabular}[c]{@{}l@{}}frappe/\\ erpnext\end{tabular}                      & 14,934                                                               & 16,700                                                              & \begin{tabular}[c]{@{}l@{}}support;python;distribution;\\ erp;accounting;crm;healthcare;\\ project-management;\\ manufacturing;frappe;\\ erpnext;procurement;\\ retail;point-of-sale;\\ hrms;asset-management\end{tabular}                                                                        \\ \hline
IT-Manufacturing & \begin{tabular}[c]{@{}l@{}}vernemq/\\ vernemq\end{tabular}                     & 1,436                                                                & 3,100                                                               & \begin{tabular}[c]{@{}l@{}}mqtt;iot;m2m;erlang;scalable;\\ messaging;pubsub;distributed;\\ message-queue;iot-middleware;\\ broker;manufacturing;\\ industrial-automation;vernemq;\\ industrial-communication;\\ industrial-iot;industry-40;\\ vernemq-users;vernemq-\\ documentation\end{tabular} \\ \hline
IT-Healthcare    & \begin{tabular}[c]{@{}l@{}}pliang279/\\ awesome-\\ multimodal-ml\end{tabular}  & 115                                                                  & 5,300                                                               & \begin{tabular}[c]{@{}l@{}}machine-learning;natural-\\ language-processing;\\ reinforcement-learning;\\ computer-vision;deep-\\ learning;robotics;healthcare;\\ reading-list;representation-\\ learning;speech-processing;\\ multimodal-learning\end{tabular}                                     \\ \hline
IT-Healthcare    & \begin{tabular}[c]{@{}l@{}}openemr/\\ openemr\end{tabular}                     & 2,399                                                                & 2,600                                                               & \begin{tabular}[c]{@{}l@{}}emr;windows;linux;php;osx;\\ sponsors;health;medical;\\ healthcare;fhir;global-health;\\ hit;ehr;international;openemr;\\ practice-management;\\ medical-information;\\ medical-informatics;medical-\\ records;proprietary-counterparts\end{tabular}                   \\ \hline
IT-Web           & \begin{tabular}[c]{@{}l@{}}donnemartin/\\ system-design-\\ primer\end{tabular} & 280                                                                  & 2,51,000                                                            & \begin{tabular}[c]{@{}l@{}}python;design;development;\\ programming;web;system;\\ design-patterns;interview;\\ web-application;webapp;\\ interview-practice;interview-\\ questions; design-system\end{tabular}                                                                                    \\ \hline
IT-Web           & \begin{tabular}[c]{@{}l@{}}TryGhost/\\ Ghost\end{tabular}                      & 6,829                                                                & 45,600                                                              & \begin{tabular}[c]{@{}l@{}}nodejs;javascript;cms;\\ blogging;journalism;ghost;\\ publishing;web-application;\\ jamstack;hacktoberfest;\\ headless-cms;creator-economy\end{tabular}                                                                                                                \\ \hline
IT-Finance       & \begin{tabular}[c]{@{}l@{}}firefly-iii/\\ firefly-iii\end{tabular}             & 5,609                                                                & 14,000                                                              & \begin{tabular}[c]{@{}l@{}}linux;docker;php;money;\\ personal-finance;finance;\\ php7;accounting;financial;\\ credit-card;budgeting;\\ cash-flow;budget;expenses;\\ finances;cashflow;\\ budgets;paycheck\end{tabular}                                                                            \\ \hline
\end{longtable}

After applying these criteria, we identified 11 top-starred projects as mentioned in Table~\ref{tab:githubrepos}. 
As listed in the table, we picked at least one project from each domain, except the IT-Government and IT-Telecommunications domains, as no projects (to the best of our knowledge) met our above acceptance criteria. 
Based on these 11 projects, we randomly selected 1,100 issues from all of the projects, i.e., almost 100 issues per repository. We tried utilising an updated version of the DistilBERT classifier\footnote{https://github.com/vedantchauhan/hcd\_collection\_analysis.git} created by~\citep{houriehhcdinapps} on GitHub issues and comments. However, it did not provide much success in the outcomes of the classification of HCDs and non-HCDs. The rationale was that the categories used in our study differ from those in~\citep{houriehhcdinapps}. Additionally, the accuracy of our updated model was significantly low due to the insufficient amount of appropriate training data. Hence, we manually classified all the issues.

\begin{figure*}
  \includegraphics[width=0.99\textwidth]{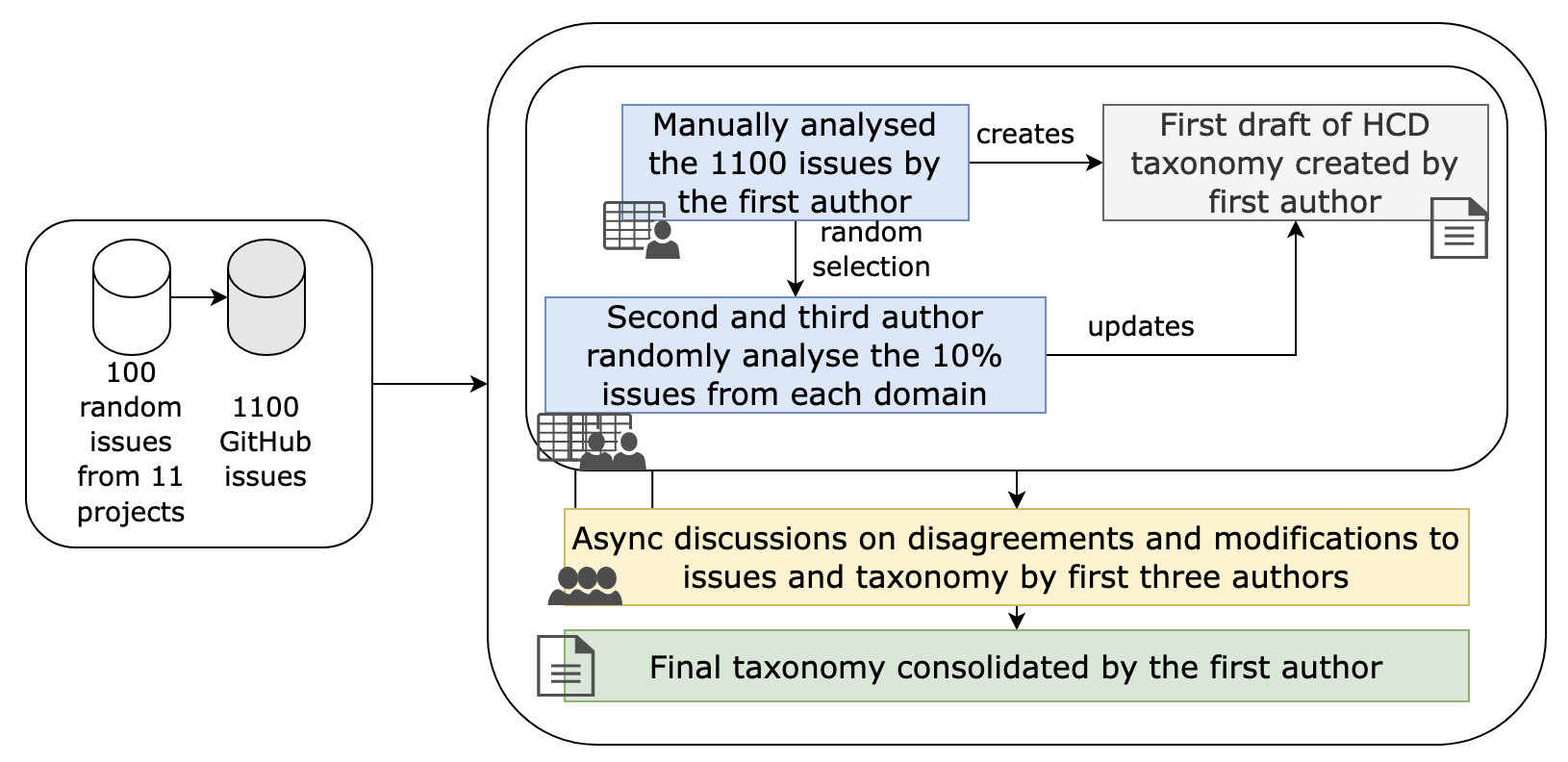}
\caption{An overview of the classification process}
\label{fig:classificationProcess}       
\end{figure*}

All 1,100 HCDs were classified into HCDs or non-HCDs. The first author classified all 1,100 issues into six key categories—five key HCD categories, i.e., functional and non-functional (usability, security, compatibility, and performance) and non-HCD. The second and third authors also categorised 10\% randomly selected issues. Disagreements, if any, were settled through deliberations among all three authors by openly elucidating the rationale behind their decisions. From the 1,100 issues, there were only 20 disagreements which were classified as non-HCD. For instance, issue no. 5783\footnote{https://github.com/openemr/openemr/issues/5783} in OpenEMR healthcare project with summary \textit{``LBF graphs using data entry dates instead of visit dates"} was disputed. The first author considered it as HCD in the functional defect category while the second and third authors considered it as normal functional defect. Therefore, it was considered as a normal functional defect. Each of the three authors kept individual spreadsheets to track their efforts. Each spreadsheet included the following information: issue text, issue type, issue number, HCD/non-HCD classification, and if the issue is HCD, one or more HCD categories it is associated with, as well as the reviewer comments. Out of 1,100 issues, there were 83 HCDs in the functional defect category. Out of those 83, there were 19 HCDs which could also fall into HCDs in the usability category. For instance, issue no. 5743\footnote{https://github.com/openemr/openemr/issues/5743} in OpenEMR healthcare project with summary \textit{``String-based route types do not save"} presents options that users are unable to save prescriptions. This issue can be considered both a functional HCD, as it pertains to the application's ability to save data, and a usability HCD, as it may result from a malfunctioning user interface button.

In parallel, the first author created a preliminary defect categorisation from the classification of issues as mentioned in Section~\ref{sec:rq1}. The defect categorisation explains the factors that can lead to HCDs in functional and non-functional (usability, security, compatibility and performance) categories. Utilising this defect categorisation, the second and third authors randomly selected 10\% of the issues in each domain category and cross-checked the findings. Based on their analysis, the defect categorisation was updated, and this process was repeated until all disagreements were resolved. There were only 20 disagreements from 1,100 issues which were classified as non-HCD. These spreadsheets are then consolidated by the first author. We cross-validated our classification findings with the 15 SE practitioners we interviewed. Each participant was given one example from each defect category to determine whether the defect belonged to a specific HCD category. The methodology for this process is detailed in Section~\ref{sec:interview}, and the results are discussed in Section~\ref{sec:rq1}. The scripts and data utilised for the defect categorisation exist in our GitHub repository\footnote{https://github.com/vedantchauhan/hcd\_collection\_analysis.git}.The process as mentioned in Figure~\ref{fig:classificationProcess} resulted in the following observations. These observations are:

\begin{itemize}
    \item Observation 1: HCD categories are not mutually exclusive, meaning that an issue can be assigned to more than one HCD category. For example, a functional defect in the ``system-design-primer" on the link excluded for documentation could also relate to usability issues if the link is broken.
    \item Observation 2: The process of labelling should not be influenced by keywords. As an illustration, we encountered numerous instances when issues lacked phrases relating to bugs, however, we did not categorise them as bugs. They could be questions related to the basic functionality of the application and vice versa. For instance, \textit{"...I am raising this question to understand the working of the application. What is the reason for running web servers as a Reverse proxy?"} - [system-design-primer].
    \item Observation 3: In our classification of 1,100 issues, only 176 were identified as HCDs. This analysis indicates that HCD concerns are infrequently addressed in open-source repositories.
\end{itemize}

\subsection{Interview Process: Design and Questionnaire}
\label{sec:interview}

\begin{figure*}
  \includegraphics[width=0.99\textwidth]{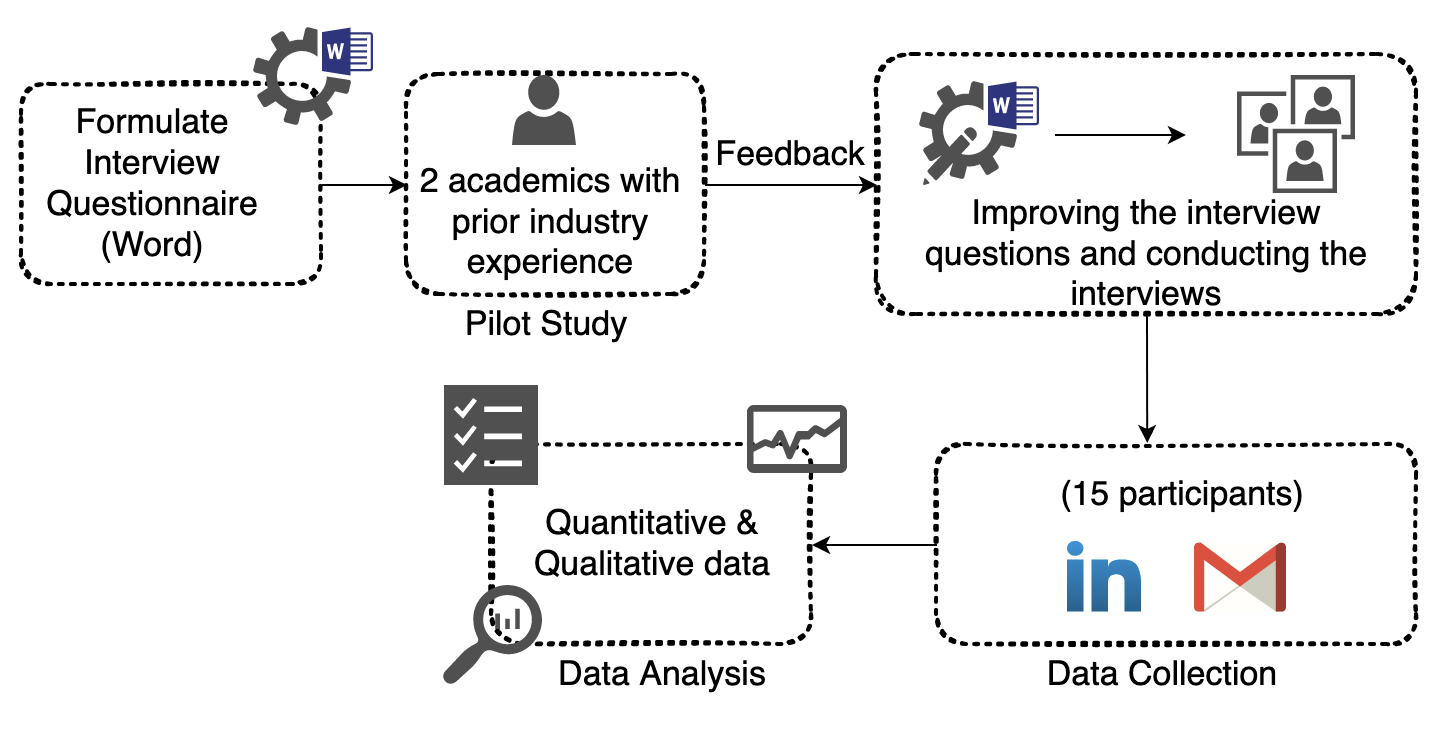}
\caption{Interview process}
\label{fig:interviewProcess}       
\end{figure*}

To validate our findings from the classification process and enhance the reporting of HCDs in defect reporting tools, we decided to conduct an interview study. Figure~\ref{fig:interviewProcess} demonstrates the process we utilised when designing interviews. The questions were formulated based on the principles outlined by Kitchenham et al. \citep{kitchenham2002preliminary,kitchenham2002principles} as well as the first author's expertise in the IT industry. Additionally, insights were gathered from various studies on defect reporting, input from industry practitioners, and considerations of human-centric aspects \citep{bettenburg2008makes, yusop2016reporting, yusop2016influences,votipka2018hackers,yusop2020revised, huynh2021improving, hidellaarachchi2021effects,chauhan2024software}. The interview consisted of a total of 22 questions, which were separated into three sections. The estimated duration for completing the interview was around 45-50 minutes. We divided our interview study into three categories: demographics, defect reporting categories and the management of HCDs in defect reporting tools. The interview questions can be summarised as follows:

\begin{itemize}
    \item \sectopic{Demographics.} The first part included the demographic questions. The interview data was anonymised, and no personally identifiable information of participants was retained. Upon completing the session, we transcribed the interview and subsequently erased the recordings.
    \item \sectopic{Defect reporting categories.} The second section contained a semi-structured set of questions to present the practitioners with examples of HCD categories, exploring the defect report in the examples and gathering practitioners' thoughts on the categorisation.
    \item \sectopic{Defect reporting structure.} The third section contained a semi-structured set of questions to continue the conversation on defect report structure, the reporter's background in the defect reports, tools' design improvement ideas, and overall management of HCDs in defect reporting tools.
\end{itemize}

Our interview questions are \textbf{available on Zenodo, please see ~\cite{anonymous_survey_and_interview}} for more details.

After designing the interview, we conducted a pilot study with two academic researchers with prior experience in the IT industry. We utilised convenience sampling for the pilot and final study for the participants who met the inclusion criteria. The criteria for the target population were current or previous experience as software developers, reporters, requirements engineers, data scientists, UI/UX specialists, and project managers who have been working or managing software systems and defect reporting tools. From the pilot study, we determined that including three GitHub examples for each defect category significantly increased the time required for the interview study. Consequently, we decided to reduce the number of examples to one per defect category to streamline the process. Based on the feedback from the pilot study, we improved the clarity and understanding of the questionnaires, improving definitions and the overall design process of the interviews. Following a successful pilot study, we proceeded to carry out a total of 15 interviews spanning from April 2023 to May 2023. We recruited participants who provided consent in our previous study to reach out for the follow-up studies. Additionally, to reduce selection bias, in the study, we utilised social media platforms such as LinkedIn to recruit participants who met our inclusion criteria as stated above. After conducting the interviews, we offered our interview participants the option to receive a 25\$ gift voucher in their respective currency from  Myers\footnote{https://www.myer.com.au/}, Coles\footnote{https://www.coles.com.au/}, iTunes\footnote{https://www.apple.com/au/itunes/}, and Amazon\footnote{https://www.amazon.com/} for dedicating their time and actively participating in the interview study. We gathered both qualitative and quantitative data by conducting interviews. We employed descriptive statistical analysis on the quantitative data using the Qualtrics platform, as shown in Section~\ref{sec:demographics}. 

We employed thematic synthesis for the qualitative data, as recommended by \citep{cruzes2011recommended}. Thematic analysis is a method that involves identifying, analysing, and reporting patterns or themes within qualitative data \citep{braun2006using}. Nevertheless, if the thematic analysis is not utilised within a fundamental conceptual framework, its explanatory capacity is limited to mere description \citep{braun2006using}. Therefore, we made the decision to employ thematic synthesis. Thematic synthesis is an extension of thematic analysis that aims to identify, explain, and summarise the themes that have been identified during the analytical process. We employed this notion to identify the repeating patterns in the qualitative data and ascertain the purpose and summary of those patterns for our overarching research inquiries. Upon doing an analysis of the qualitative data, we successfully distinguished distinct text segments within the dataset. Subsequently, we employed an open coding technique to assign labels to the text, with the aim of constructing ideas, concepts, and overarching themes~\citep{khandkar2009open}.

For instance, in the interviews, we asked participants, ``Do different types of HCDs and domains impact the severity and priority of the issue?". Based on the fifteen responses, we received lower-order codes such as impact importance for human-centric issues, severity and priority of the defects, impact and severity for the user and teams, impact for different users, and how to identify the priority of the issues. Based on these low-order codes, higher-order themes such as the impact's importance to the user, practitioners' understanding of severity and priority, and identifying the impact on the user are generated. Similarly, we used the open-coding technique for other open-ended questions by constant comparison, creating lower-order codes and merging them into higher-order themes.

\subsection{Participant Demographics}
\label{sec:demographics}

Table~\ref{table:interviewdemographics} shows the demographics of our 15 recruited interview participants. We received most responses from Australia (11), while the remaining participants were from Canada and the USA. All fifteen participants had a university degree in Computer Science, Information Technology, and Clinical Physics and had previous or current experience in using defect reporting tools. Out of 15 participants, 12 participants use the defect reporting tools sometimes, always, and often; only 3 participants rarely use the defect reporting tools in their day-to-day activities. In our results reporting, interview participants are referred to by \emph{IP-x}. We considered both developers and reporters of our interview as practitioners.

{\footnotesize
\begin{longtable}{@{}ll@{}}
\caption{Interview participants' demographics information}

\label{table:interviewdemographics}\\

\toprule
\textbf{Demographic Information of the participants}      & \textbf{Overall} \\* \midrule
\endhead
\bottomrule
\endfoot
\endlastfoot
\textbf{Countries of the participants}                    &                  \\* \midrule
Australia                                                 & 11           \\
USA                                                       & 3             \\
Canada                                                    & 1            \\* \midrule
\textbf{Age ranges of the participants}                   &                  \\* \midrule
30-34                                                     & 12             \\
35-39, 40-50, and Prefer not to say                       & 3             \\* \midrule
\textbf{Gender of the participants}                       &                  \\* \midrule
Male                                                      & 10           \\
Female                                                    &  4          \\
Prefer not to say                                         & 1            \\* \midrule
\textbf{Highest Education}                                &                  \\* \midrule
Master's degree                                           & 13           \\
Doctorate degree                                          & 1            \\
Bachelor's degree                                         & 1            \\* \midrule
\textbf{Education information of the participants}        &                  \\* \midrule
Information Technology                                    & 8           \\
Computer Science                                          & 6             \\
Clinical Physics                                          & 1           \\* \midrule
\textbf{Job roles of the participants}                    &                  \\* \midrule
Software Engineer/Software Developer/Programmer           & 7           \\
Test Engineer/Tester/Quality Assurance                    & 5           \\
Consultant                                                & 1            \\
Data Scientist                                            & 1            \\
Data Engineer                                             & 1            \\* \midrule
\textbf{Total Experience of participants in IT industry}  &                  \\* \midrule
3-5 years                                                 & 3          \\
6-8 years                                                 & 8          \\
9+ years                                                  & 4           \\* \midrule
\textbf{Field of Work}                                    &                  \\* \midrule
IT                                                        & 8           \\
IT-Finance                                                & 3             \\
IT-Healthcare                                             & 2           \\
IT-Government                                             & 1           \\
IT-Construction                                           & 1            \\* \midrule
\textbf{Total Experience of using defect reporting tools} &                  \\* \midrule
0.1-2 years                                               & 3             \\
3-5 years                                                 & 8           \\
6-8 years                                                 & 3             \\
9+ years                                                  & 1            \\* \midrule
\textbf{How often do you use defect reporting tools}      & \textbf{}        \\* \midrule
Always                                                    & 5           \\
Often                                                     & 2           \\
Sometimes                                                 & 5           \\
Rarely                                                    & 3             \\* \bottomrule
\end{longtable}
}


\section{Results}
\label{sec:results}

\subsection{RQ1 Results - Classification and defect categorisation of HCDs and non-HCDs}
\label{sec:rq1}

\begin{figure}
\centering
  \includegraphics[width=\textwidth]{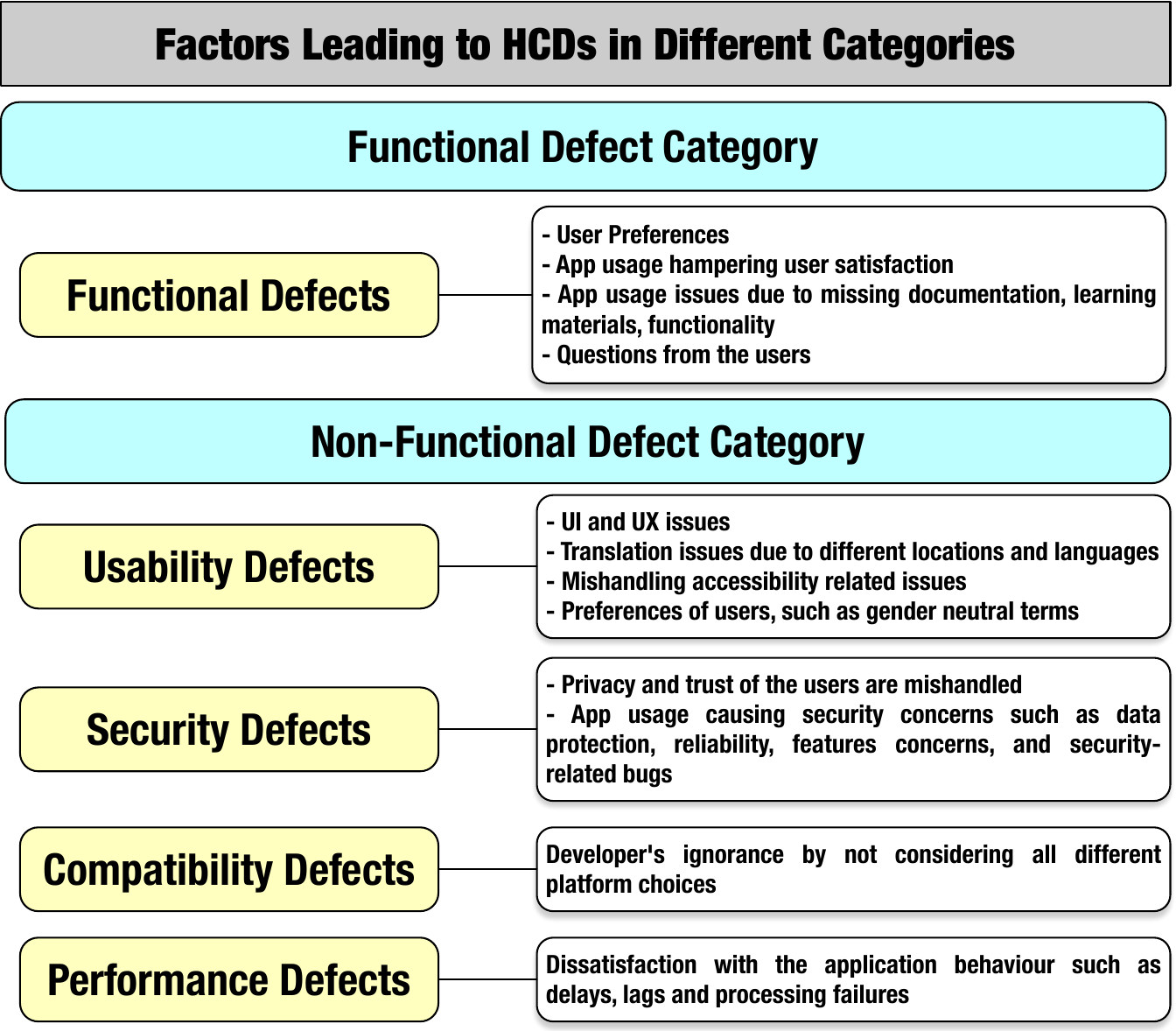}
\caption{Defect categorisation of HCDs in different defect categories}
\label{fig:hcdtaxonomy}       
\end{figure}

\begin{figure}
\centering
  \includegraphics[width=\textwidth]{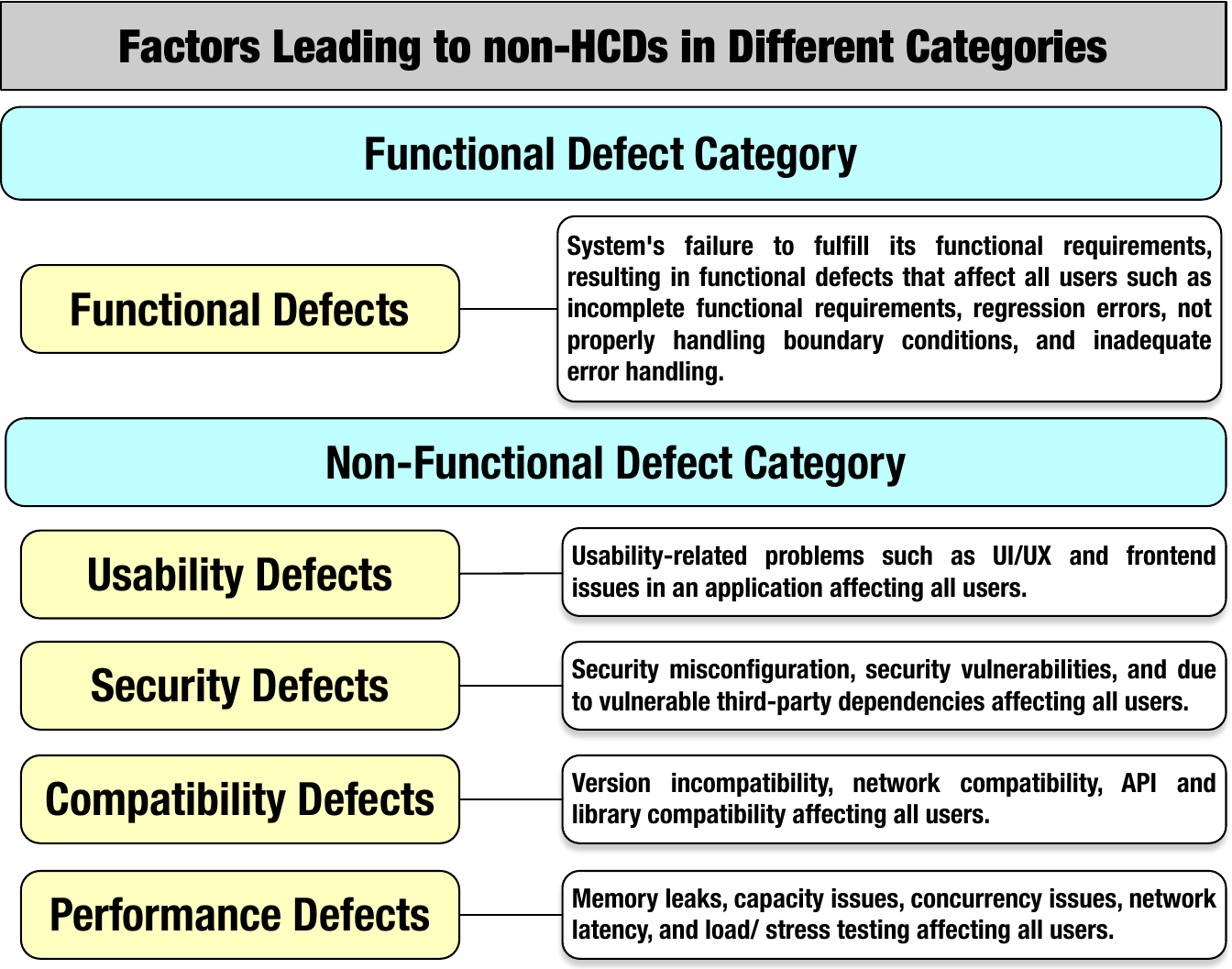}
\caption{Defect categorisation of non-HCDs in different defect categories}
\label{fig:nonhcdtaxonomy}       
\end{figure}

In our previous study~\citep{chauhan2024software}, we asked practitioners to highlight HCDs in defect categories, such as functional, performance, security,
usability, and compatibility~\citep{matrinnfr, eckhardt2016non}. Based on the practitioner's responses, it turns out that HCDs are frequently reported in the usability category followed by functional and compatibility and concludes with
the performance and security defect categories. This RQ provides triangulation of the results and provides the analysis of GitHub issues to validate the claims provided by practitioners. Based on our analysis, we created a preliminary defect categorisation of HCDs as mentioned in the above-reported defect categories. For instance, we provide factors that can lead to an HCD in a functional defect category. We note that these factors are not exhaustive; we categorise them based on our analysis of 1,100 issues, and a more exhaustive analysis can potentially lead to more categories of HCDs. Figs.~\ref{fig:hcdtaxonomy} and~\ref{fig:nonhcdtaxonomy} provide a detailed diagram of the HCD and non-HCD defect categorisation based on our analysis of 1,100 Github issues. Additionally, we discussed an example from each of the five key categories with 15 practitioners to provide validation on the HCDs in the following categories. 

\subsubsection{Functional HCDs}

HCDs can occur in functional defects when a user interacts with the application for several reasons.
Defects that can arise from a user's point of view i.e., preferences for using the application. For instance, a user raised an issue about adding the link to understand the concepts provided by the application, \textit{``...add the Youtube video link which can help in understanding denormalisation with an analogy in the application..."} - [system-design-primer].\\
Defects can occur due to the usage of the application, i.e., hampering the satisfaction of the user while using the application due to page load issues, access issues, missing free signups, etc. Additionally, due to missing documentation, learning materials, and missing functionality. For instance, a user raised an issue, \textit{``I wanted to use the library and discovered the installation section of the documentation. The only documented way is conda, even though the library is  published and up-to-date on Pipy. It didn't take me long to figure it out but that was still a missing information. I would like the documentation to be up-to-date."} - [osmnx]. \\
Issues that can be questions and not necessarily defects but can lead to functionality missing and causing user hindrance. For example, a question raised by a user, \textit{``...I am raising this question to understand the working of the application. What is the reason for running web servers as a Reverse proxy?"} - [system-design-primer].

Based on discussions with practitioners, we reviewed issue no. 15426\footnote{https://github.com/TryGhost/Ghost/issues/15426} with 15 practitioners. In our classification, this defect was categorised as a functional issue based on application usage. The issue, summarised as \textit{``Can't login with self domain mail,"} involves a user being unable to log in with their existing email, although login with Gmail works. Among the 15 software engineering practitioners, 7 identified this defect as an HCD issue, 3 as a possible HCD issue, and 5 as a non-HCD issue.

Practitioners who classified this defect as an HCD issue provided reasoning such as, \textit{``The developer didn't consider different domains and may have added certain checks, which skipped or did not validate these kinds of domains, thus triggering exception scenarios in the code and causing an HCD issue for the user."} On the other hand, those who categorised it as a non-HCD issue argued, \textit{``This usually happens when SMTP parameters are not set properly. It is likely that they recently migrated their email-sending software to a different domain or API, which might have led to this technical functional failure."} From the findings of the interview study with practitioners, we continued to classify this HCD under the functional defect category.

\subsubsection{Usability HCDs}

HCDs can occur in usability defects when a user interacts with the application for several reasons.
Defects that occur due to User Interface (UI) \& User Experience (UX) related issues. For example, a user raised an issue, \textit{``...when you zoom out, the screen shows the grey square boxes. Otherwise, the application works as expected."} - [pixijs].\\
Defects that occur to translation such as location and language-related issues. For instance, a user raised an issue due to foreign currency problems, \textit{``I've tried to activate a foreign currency (by default I use EUR, and I want to add US Dollar), but when I try to add on a new transaction, not appear in the list."} - [firefly-iii].
Defects that occur due to mishandling of accessibility issues. For instance, \textit{``...I couldn't use the keyboard shortcut for this feature. Can you look into this?"} - [firefly-iii].\\
Defects that occur due to user preferences such as the usage of gender-neutral terms in the application. For e.g.,  \textit{``We noticed your repository had a few instances of gendered language. You can learn more about this project and why gender-neutral language matters at inclusivecoding.wixsite.com. These are not always perfect, but we hope they will assist maintainers in finding and fixing issues. Can you switch to gender-neutral terms?"} - [system-design-primer].

We reviewed issue no. 5771\footnote{https://github.com/openemr/openemr/issues/5771} with 15 practitioners. In our classification, this defect was categorised as a usability issue based on UI \& UX. The issue summarised as \textit{``OpenEMR video tutorials and work flows,"} where the application lacks updated video tutorials and does not provide an optimal user experience (UX) for the user. Among the 15 SE practitioners, 14 identified this defect as an HCD issue and one as a non-HCD issue.

Practitioners who classified this defect as an HCD issue provided reasoning such as, \textit{``If your application provides video tutorials of older versions, and the steps in the older version are no longer applicable, then the user might just get confused as to something is wrong."} On the other hand, the practitioner who categorised it as a non-HCD issue argued, \textit{``When using an application, there are numerous instances in the market where documentation or tutorials are lacking, leading to a smaller customer base. Developers tend to overlook defects in less popular applications, unlike those from prominent organisations such as Uber or Lyft. While a defect preventing all users from accessing video content would be considered a functional issue, a similar problem affecting only one user is often deemed insignificant by the organisation."} From the findings of the interview study with practitioners, we continued to classify this HCD under the usability defect category.

\subsubsection{Security HCDs}

HCDs can occur in security defects when a user interacts with the application for several reasons.
Defects that can cause concerns to the privacy and trust of the users. For instance, the user's privacy is affected due to auto-approval settings in the application, \textit{``Currently a public app is auto-approved in OpenEMR, whereas all confidential apps are required to be manually approved. We should change up the logic so that if a confidential app requests only patient scopes and not user or system scopes the app is auto-approved. We should also add a global setting if a user wishes to change the logic to require manual approval of all apps."} - [openemr].\\
Defects that can cause security concerns during usage of the application such as data protection, reliability, features concerns, and security-related bugs. For example, a user misread documentation to provide short keys, \textit{``I was trying to set up 2FA, but after submitting the first code, I get this Secret key is too short. Must be at least 16 base32 characters error."} - [firefly-iii].

Based on discussions with practitioners, we reviewed issue no. 6402\footnote{https://github.com/firefly-iii/firefly-iii/issues/6402} with 15 SE practitioners. In our classification, this defect was categorised as a security issue based on application usage due to feature bugginess. The issue, summarised as \textit{``2FA Setup Secret key is too short. Must be at least 16 base32 characters,"} where the 2FA security key is not working as expected and giving user unexpected results or errors. Among the 15 software engineering practitioners, 7 identified this defect as an HCD issue, 6 as a possible HCD issue, and 2 as a non-HCD issue.

Practitioners who classified this defect as an HCD issue provided reasoning such as, \textit{``When the user is trying to set up the two factor authentication in the UI. The application can provide details to user such as the password is too short, rather than sending them emails with such information because not everyone is going to be tech savvy person to understand such issues."} On the other hand, those who categorised it as a non-HCD issue claimed, \textit{``Regardless of the six-digit code entered, the error message states that the secret key is too short. This issue is not human-centered because the user is entering the provided code, and even trying different six-digit codes, yet the application consistently fails to function as intended. This problem affects all users, irrespective of the code they enter."} From the findings of the interview study with practitioners, we continued to classify this HCD under the security defect category.

\subsubsection{Compatibility HCDs}

HCDs can occur in compatibility defects when a user interacts with the application due to the following reasoning.\\
Defects that occur due to developer’s ignorance by not considering all different platform choices where users can view the content such as mobiles, tablets, desktops, etc. \citep{houriehhcdinapps}. For instance, \textit{``...graphics repeats texture on desktop but not on mobile."} - [pixijs].

Based on discussions with practitioners, we reviewed issue no. 8543\footnote{https://github.com/pixijs/pixijs/issues/8543} with 15 SE practitioners. In our classification, this defect was categorised as a compatibility issue based on developer's ignorance to not consider all platforms. The issue, summarised as \textit{``Graphics.beginTextureFill repeats texture on desktop but not on mobile,"} where rendering a graphics texture works on desktop but does not on mobile. Among the 15 software engineering practitioners, 9 identified this defect as an HCD issue, 2 as a possible HCD issue, and 4 as a non-HCD issue.

Practitioners who classified this defect as an HCD issue provided reasoning such as, \textit{``As users, regardless of the platform we are using—whether it is a desktop, mobile device, or any other user interface—the experience should be consistent and uniform.."} On the other hand, those who categorised it as a non-HCD issue claimed, \textit{``This issue does not affect a specific group of users; rather, it affects all users equally. If the application behaves in this manner, it impacts every user, not just a particular subset. Therefore, it should not be classified as a human-centered defect."} From the findings of the interview study with practitioners, we continued to classify this HCD under the compatibility defect category.

\subsubsection{Performance HCDs}

HCDs can occur in performance defects when a user interacts with the application due to the following reasoning.\\
Defects that arise due to the user experiencing dissatisfaction with the application behaviour such as delays, lags, processing failures, and so on. For instance, \textit{``I am unable to import large CSV files greater than 100 MB."} - [firefly-iii].

Based on discussions with practitioners, we reviewed issue no. 6337\footnote{https://github.com/firefly-iii/firefly-iii/issues/6337} with 15 SE practitioners. In our classification, this defect was categorised as a performance issue based on user's dissatisfaction on application behaviour. The issue, summarised as \textit{``Unable to import large csv,"} where the application is not allowing the user to upload large CSV files and showing a  memory limit error. Among the 15 software engineering practitioners, 10 identified this defect as an HCD issue, 4 as a possible HCD issue, and 1 as a non-HCD issue.

Practitioners who classified this defect as an HCD issue provided reasoning such as, \textit{``I believe the application does not permit the upload of larger files. Although this information may not be readily available, the issue specifically affects users attempting to upload large CSV files. Therefore, I consider it a human-centered defect."} On the other hand, those who categorised it as a non-HCD issue claimed, \textit{``The presence of a large and extensive file is pushing the application to its limits. However, I do not believe this issue falls under the category of human-centered defects. I think human-centered defects emphasises the interaction of individuals with the application itself, rather than the specific features or functionalities they are utilising."} From the findings of the interview study with practitioners, we continued to classify this HCD under the performance defect category.

Based on our analysis, we created a distinguishing characteristic for non-HCDs i.e. ``normal technical defects" as follows from the 1,100 GitHub issues:

\sectopic{Functional defects.} These problems typically arise from the system's failure to fulfil its functional requirements, resulting in functional defects that affect all users. These defects can be due to incomplete functional requirements, regression errors, not properly handling boundary conditions, and inadequate error handling. For instance, an issue raised by the user on boundary conditions, \textit{``...the whole attribution data from API is missing the attribution type value, we cannot set the whole object as null now as that hides the referrer information. Please update attribution data in the API to include missing information for this edge condition."} - [ghost].

\sectopic{Usability defects.} These defects arise due to usability-related problems such as UI/UX and frontend issues in an application affecting all users. For instance, \textit{``...all our new users are experiencing broken layout of the home page. The error resulted due to new logins created and using the application. Can you look into this?"} - [firefly-iii].

\sectopic{Security defects.} These defects arise due to security-related problems in an application affecting all users. These defects can be due to security misconfiguration, security vulnerabilities, and vulnerable third-party dependencies. For example, a security misconfiguration resulting in the TLS handshake failure, \textit{``TLS client: A state cipher received SERVER ALERT: Fatal - Handshake Failure on a vernemq cluster."} - [vernemq].

\sectopic{Compatibility defects.} These defects arise due to compatibility-related problems in an application affecting all users. These defects can be due to version incompatibility, network compatibility, API and library compatibility. For instance, a technical issue on Podman and Docker containers incompatibility, \textit{``Podman is a similar platform to Docker, yet mainly runs on Linux and is used in enterprise solutions. Red Hat Enterprise Linux utilises this. Unfortunately, the docker-compose files don't seem fully compatible with it."} - [openemr].

\sectopic{Performance defects.} These defects arise due to performance-related problems in an application affecting all users. These defects can be due to memory leaks, capacity issues, concurrency issues, network latency, and load/stress testing. E.g., \textit{``VerneMQ connections not closing causing it to crash."} - [vernemq].

Table~\ref{table:classhcds} provides final details for our classification of HCDs into different defect reporting categories based on our analysis and from the discussion with practitioners. Out of the 1,100 issues, 176 were HCDs and the remaining 924 were non-HCDs based on our factors. Based on the results, we identified that functional defect category had more HCDs followed by usability, performance, security and compatibility. The outcome differs from what practitioners asserted in our prior study. However, both Additionally, IT-Manufacturing had the most HCDs followed by IT-Web, IT-Healthcare, IT-Financial, IT-Gaming, and IT-Spatial.
\begin{table}[]
\centering
\caption{Classification of HCDs}
\label{table:classhcds}
\begin{tabular}{|l|lllll|l|l|}
\hline
\multirow{2}{*}{\textbf{Domain}} & \multicolumn{5}{c|}{\textbf{Defect Categories}}                                                                                                                                                  & \multirow{2}{*}{\begin{tabular}[c]{@{}l@{}}{\textbf{Total}} \\ {\textbf{HCDs}}\end{tabular}} & \multirow{2}{*}{\begin{tabular}[c]{@{}l@{}}{\textbf{Total}} \\ {\textbf{Non-HCDs}}\end{tabular}} \\ \cline{2-6}
                                 & \multicolumn{1}{l|}{\rotatebox[origin=c]{90}{\textbf{Functional}}} & \multicolumn{1}{l|}{\rotatebox[origin=c]{90}{\textbf{Usability}}} & \multicolumn{1}{l|}{\rotatebox[origin=c]{90}{\textbf{Security}}} & \multicolumn{1}{l|}{\rotatebox[origin=c]{90}{\textbf{ Compatibility }}} & \rotatebox[origin=c]{90}{\textbf{Performance}} &                                                                       &                                                                           \\ \hline
IT-Healthcare                    & \multicolumn{1}{l|}{23}                  & \multicolumn{1}{l|}{12}                 & \multicolumn{1}{l|}{2}                 & \multicolumn{1}{l|}{0}                      & 0                    & 37                                                                    & 158                                                                       \\ \hline
IT-Web                           & \multicolumn{1}{l|}{14}                  & \multicolumn{1}{l|}{25}                 & \multicolumn{1}{l|}{0}                 & \multicolumn{1}{l|}{0}                      & 0                    & 39                                                                    & 161                                                                       \\ \hline
IT-Spatial                       & \multicolumn{1}{l|}{9}                   & \multicolumn{1}{l|}{3}                  & \multicolumn{1}{l|}{0}                 & \multicolumn{1}{l|}{0}                      & 4                    & 16                                                                    & 184                                                                       \\ \hline
IT-Manufacturing                 & \multicolumn{1}{l|}{19}                  & \multicolumn{1}{l|}{17}                 & \multicolumn{1}{l|}{0}                 & \multicolumn{1}{l|}{0}                      & 5                    & 41                                                                    & 159                                                                       \\ \hline
IT-Financial                     & \multicolumn{1}{l|}{13}                  & \multicolumn{1}{l|}{8}                  & \multicolumn{1}{l|}{2}                 & \multicolumn{1}{l|}{1}                      & 1                    & 25                                                                    & 80                                                                       \\ \hline
IT-Gaming                        & \multicolumn{1}{l|}{5}                   & \multicolumn{1}{l|}{12}                 & \multicolumn{1}{l|}{0}                 & \multicolumn{1}{l|}{1}                      & 0                    & 18                                                                    & 182                                                                       \\ \hline
\textbf{Total}                   & \multicolumn{1}{l|}{\textbf{83}}         & \multicolumn{1}{l|}{\textbf{77}}        & \multicolumn{1}{l|}{\textbf{4}}        & \multicolumn{1}{l|}{\textbf{2}}             & \textbf{10}          & \textbf{176}                                                          & \textbf{924}                                                              \\ \hline
\end{tabular}
\end{table}

\begin{tcolorbox}[arc=0mm,width=\columnwidth,
                  top=1mm,left=1mm,  right=1mm, bottom=1mm,
                  boxrule=1pt] 
\textbf{RQ1 Key findings.} 
\begin{itemize}
    \item Out of 1,100 GitHub issues, 176 were classified as HCDs and the remaining as non-HCDs as mentioned in Table~\ref{table:classhcds}.
\item Functional defects had more HCDs followed by usability, performance, security and compatibility. This order is different from what practitioners claimed in \citep{chauhan2024software}.
\item IT-Manufacturing had the most HCDs followed by IT-Web, IT-Healthcare, IT-Financial, IT-Gaming, and IT-Spatial.
\item Based on our analysis of 1,100 GitHub issues, we created a preliminary defect categorisation for HCDs and non-HCDs in the functional and non-functional defect categories.

\end{itemize}
\end{tcolorbox}

\subsection{RQ2 Results - Existing defect reporting process}\label{sec:rq2}

Existing defect reporting tools have a generic reporting structure consisting of description, expected behaviour, observed behaviour, steps to reproduce, attachments, severity, priority, impact and product-related information as shown in Figure~\ref{fig:defectrepfields}. This RQ highlights the existing defect reporting process and its association with HCDs. Based on our discussions with the practitioners and our analysis we had three key themes, as discussed below:

\begin{figure*}
  \centering
\includegraphics[width=0.69\textwidth]{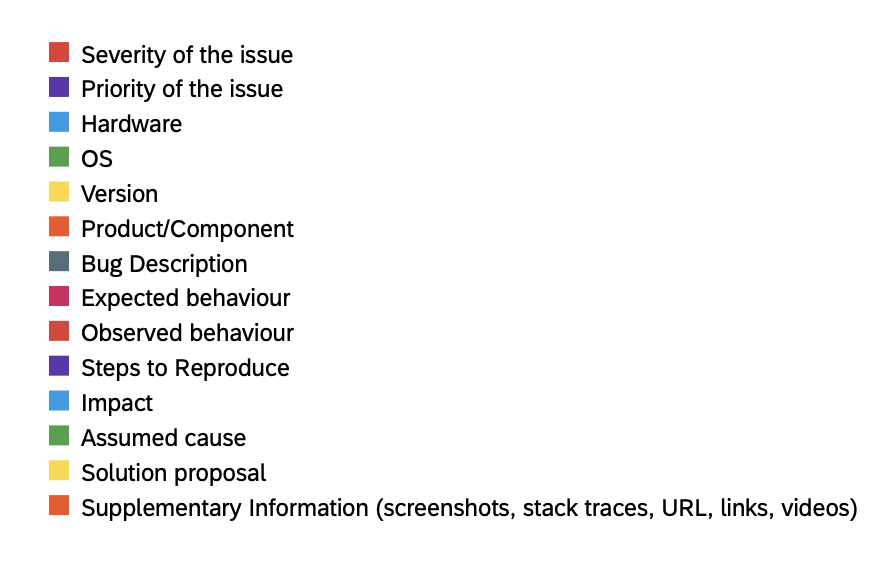}
\caption{Defect reporting fields}
\label{fig:defectrepfields}       
\end{figure*}

\subsubsection{Theme 1 - Existing reporting structure in different domains and categories of defects.}
12 out of the 15 interviewees suggested that reporting structure varies for different domains and categories of defects (such as functional and non-functional defects). For instance, IP-10 claimed, \textit{``Reporting structure should vary because not all the questions will be relevant in all the domains, and it will be easier for developers to understand if the reporting tool is customised for that particular domain."} Most practitioners claim these domains and defect categories have different severity and priorities. IP-2 highlighted, \textit{``I worked in various domains, such as in finance or banking all the issues are of the highest severity. However, in healthcare, the severity doesn't matter, the high-severity incidents in other domains have medium or low severity in healthcare. Similarly, the case with the manufacturing domain. Severity and priority are different for different domains. It depends on the defect, domain, and the category of the issue."}

According to some practitioners, most of the defect reporting fields as shown in Fig.~\ref{fig:defectrepfields}, will be similar in all the domains with slight changes in the fields such as transaction-related data in IT-Finance or patient information in IT-Healthcare. For instance, IP-15 suggested, \textit{``In the healthcare domain, bug description, steps to reproduce, hardware, OS, expected and observed behaviour will be standard across all the domains. However, a solution proposal would be different because the nurses and the doctors using the application would be able to better explain the problem related to patients rather than the developers developing it."} - [IP-15]. Another participant (IP-12) claimed, \textit{``...in finance, adding information related to transactions such as nature of transaction i.e. credit or debit and geographic location i.e. domestic or international can really provide depth to a defect."}

To accommodate HCDs in different domains and categories, human-centric experts can help in collecting the information suitable for different domains. Additionally, automation and proper tagging of the issues can also reduce the workload on development teams to identify HCDs for different domains. For example, IP-4 stated, \textit{``Automation and labelling could help prioritise human-centric issues in different domains."} Another participant (IP-5) suggested, \textit{``Experts who have knowledge and domain expertise in human-centric areas can certainly elevate the capturing of such issues in different domains via defect reporting tools."}

\subsubsection{Theme 2 - Severity, priority, and impact of defects.}

Practitioners highlight that severity is usually provided by the end-users; however, priority is always decided by the development teams based on the severity and impact of the issue. For instance, IP-10 claimed, \textit{``Severity of the issue is based on user experience. But the priority is what developers or teams need to manage based on the issue because if the issue relates to the payment system then it would be considered as high priority even though the user has added it as low severity."} The impact of the issue is also considered important from a human-centric perspective because impact provides the user's perspective or frustration with the situation they are facing. Hence, severity in association with the impact can provide the user's perception with clarity. For example, IP-15 stated, \textit{``If a customer is going through the effort of filling a bug, it means it is severe for them. In order to truly quantify an issue for a customer, the impact and severity can explain the user's frustration much better."} Another participant (IP-14), highlighted, \textit{``Severity in conjunction with impact can help in understanding the issue."}

The impact for each user is subjective. In order to consolidate the impact for different users, the development team can identify ``how many users are impacted by the issue?". For instance, IP-12 suggested, \textit{``If a defect is a business use case and how many users are affected by that defect? can help in determining the impact of the defect."} Another criterion to identify the impact of the issue is the time sensitivity of the issue or the impact caused by the issue on the business in terms of revenue or public image. For example, a payment information leak or a kidney dialysis machine being offline are high-sensitivity issues that should be prioritised first. IP-4 stated, \textit{``...if kidney dialysis machine is offline right now then this problem can definitely be on high priority due to the time-sensitive nature of the issue which can cause problems to multiple patients."} - [IP-5]. Another participant highlighted, \textit{``Project managers and teams identify the correlation between the app impact versus the company revenue getting affected due to the issue, then they prioritise the issue because it is impacting business image."}

\subsubsection{Theme 3 - Resolution of defects.}

Due to the subjective nature of HCDs, the resolution of defects could create conflict where one user can claim their defect holds greater significance. 13 practitioners stated the number of users impacted by the defect could be one factor in determining the priority of the defect. For instance, IP-10 claimed, \textit{``The easiest way is to see how many times you have received the same error. If the same error has been reported by many users of different demographics, then, of course, it is an error worth pursuing."} Additionally, practitioners stated to collect more information from the users to find quick solutions as well as concrete testing of the application to avoid problems in the first place. For instance, IP-5 claimed , \textit{``... in general as much testing as possible. It's not always possible but the teams have to include everyone in the testing so we don't exclude anyone based on human-centric aspects."}

\begin{tcolorbox}[arc=0mm,width=\columnwidth,
                  top=1mm,left=1mm,  right=1mm, bottom=1mm,
                  boxrule=1pt] 
\textbf{RQ2 Key findings.} 
\begin{itemize}
    \item Most practitioners suggested that the reporting structure should be different for different domains and categories of defects. Human-centric experts, proper tagging, and automation can help collect information suitable for different domains to accommodate HCDs in different domains and categories.
\item Severity in association with the impact can provide the user’s perception with clarity.
\item To identify the impact of the issue is the time sensitivity of the issue or the impact caused by the issue on the business in terms of revenue or public image. 
\item Most practitioners stated the number of users impacted by the defect could be one factor in determining the priority of the defect.
\end{itemize}
\end{tcolorbox}

\subsection{RQ3 Results - Does reporter's background help in capturing HCDs}
\label{sec:rq3}

We define a reporter's background as the human characteristics of the reporter such as age, gender, accessibility requirements, educational background, and technical proficiency. This research question aims to determine the effectiveness of using a reporter's background information in the defect-reporting process for resolving HCDs. 12 out of 15 practitioners suggested that capturing a reporter's background information can be useful to help developers empathise with the end-user and understand the issue from their perspective. Additionally, collecting such information can create healthy discussions among development teams while solving issues. For instance, IP-13 stated, \textit{``From the developer's point of view, it is definitely helpful because they would think about missing details such as why didn't user add this? They will empathise with the user and can ask users more about their difficulty to solve the problem."} 

Practitioners also highlighted that sometimes developers just want to solve the issues and only concern themselves with the information relevant to them to solve the issues. However, collecting a reporter's background information can still be useful because it can help product managers or owners understand their target audience. For instance, IP-3 suggested, \textit{``Assume a product with 1000 users and contains human-centric defects in physical disability which is generally an outlier. So if we assume 1\% of users are physically disabled, which is 10 people of the customer base. Out of those 10 people, if 5 are raising similar defects, then 50\% of your physically disabled people are having issues. In that scenario, changing the outlook that rather than looking at your entire user base, you look at the bucket of physically disabled and identify at least 50\% of them having similar issues. A product manager of the team can then decide to leverage that information, and try to solve that leading to a better experience for those five people in that category."}

Practitioners warn that collecting a reporter's background information can lead to privacy issues. Therefore, collecting information voluntarily and managing it sensitively after collection is important. For instance, IP-3 stated, \textit{``If you want to collect this information then the users should voluntarily provide this information. This kind of information can raise privacy concerns for the user."} IP-5 also emphasised, \textit{``I think the information about age, gender, disability, etc. user is providing, you need to be sensitive and careful about how you collect it. As you know this Personal Identifiable Information (PII) and users will have concern while providing such data."}

\begin{figure*}
  \centering  \includegraphics[width=0.59\textwidth]{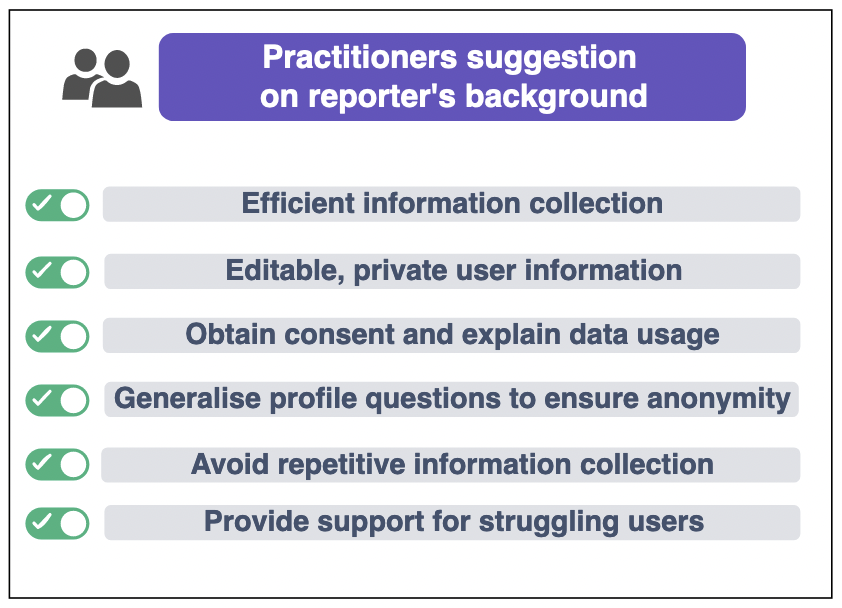}
\caption{Practices suggested by practitioners for reporter's background}
\label{fig:rbpractices}       
\end{figure*}

Practitioners suggest the following options while collecting the reporter's background information as shown in the figure~\ref{fig:rbpractices}: (i) information collection should not be time consuming to the user, (ii) users should be given the ability to modify their personal information while also ensuring that their personal information remains private and is not disclosed to others, (iii) before obtaining any sensitive information, it is imperative to obtain the user's consent. Additionally, it is essential to provide the user with comprehensive information regarding the purpose and utilisation of their data, (iv) questions presented in a profile should be generalised, ensuring that the user's identity remains anonymous and their privacy is not compromised, (v) collect the information once, so no repetitive tasks are performed by the users, and, (vi) customer support should be provided to struggling end-users. For instance, IP-10 proposed, \textit{``Three things from my perspective, get the consent of the user, provide all the questions as optional by anonymising things, and provide users why they should provide information and how it will help?"}. IP-2 claimed, \textit{``I think customer service should be provided to the people with special needs struggling by asking what issues they are facing and how those issues can be resolved. Customer reporting should be prioritised by keeping things simple and less time-consuming."}

However, 3 out of 15 practitioners claim that collecting a reporter's background information can create bias and there should be a value proposition for collecting such information. According to practitioners' perspective, the application will inevitably include defects that users have to cope with. For instance, IP-5 highlighted, \textit{``I understand why you're doing the background collection and asking people to become the subject of the data. They are giving away their private information and therefore there needs to be a value proposition in return, you need to be able to solve their problems, and you need to do it efficiently. Additionally, you need to explain in great detail why you're doing this and why they need to upload this information. And if you don't get the defect right, your value proposition fails and it will give a massive backlash from the users."} - [IP-5]. The same participant then stated, \textit{``Your end-user has to internalise the fact that errors are a part of life and people are not going to be happy if something doesn't work. It's annoying that it doesn't work and that you've got errors coming back. Consider Windows updates, we get large amounts of updates from Windows regularly because they didn't code it correctly. But we all just accept that software has defects and we have to deal with it."} - [IP-5].


\begin{tcolorbox}[%
       arc=0mm,width=\columnwidth,
                  top=1mm,left=1mm,  right=1mm, bottom=1mm,
                  boxrule=1pt]
    \textbf{RQ3 Key findings.} 

\begin{itemize}
        \item Most practitioners suggested that capturing a reporter’s background information can be useful to help developers empathise with the end-user and understand the issue from their perspective.
\item Practitioners warn that collecting a reporter’s background information can lead to privacy issues. Therefore, it is important to collect information voluntarily and manage information sensitively after collection.
\item Practitioners suggested various practices such as consent of the users, usage of the data collected, and customer support to follow while collecting the reporter's background. 
\item Some practitioners claim collecting a reporter’s background information can create bias and there should be a value proposition for collecting such information. According to practitioners’ perspective, the application will inevitably include defects that users have to cope with.
\end{itemize}
\end{tcolorbox}

\subsection{RQ4 Results - Practices for defect reporting tools to capture HCDs}
\label{sec:rq4}

Based on the practitioners'
suggestions, RQ4 provides details of the practices required by the defect reporting tools to capture HCDs efficiently (Fig.~\ref{fig:defectpractices}). Based on the discussions with practitioners, the following themes emerged:

\begin{figure*}
  \centering  \includegraphics[width=0.9\textwidth]{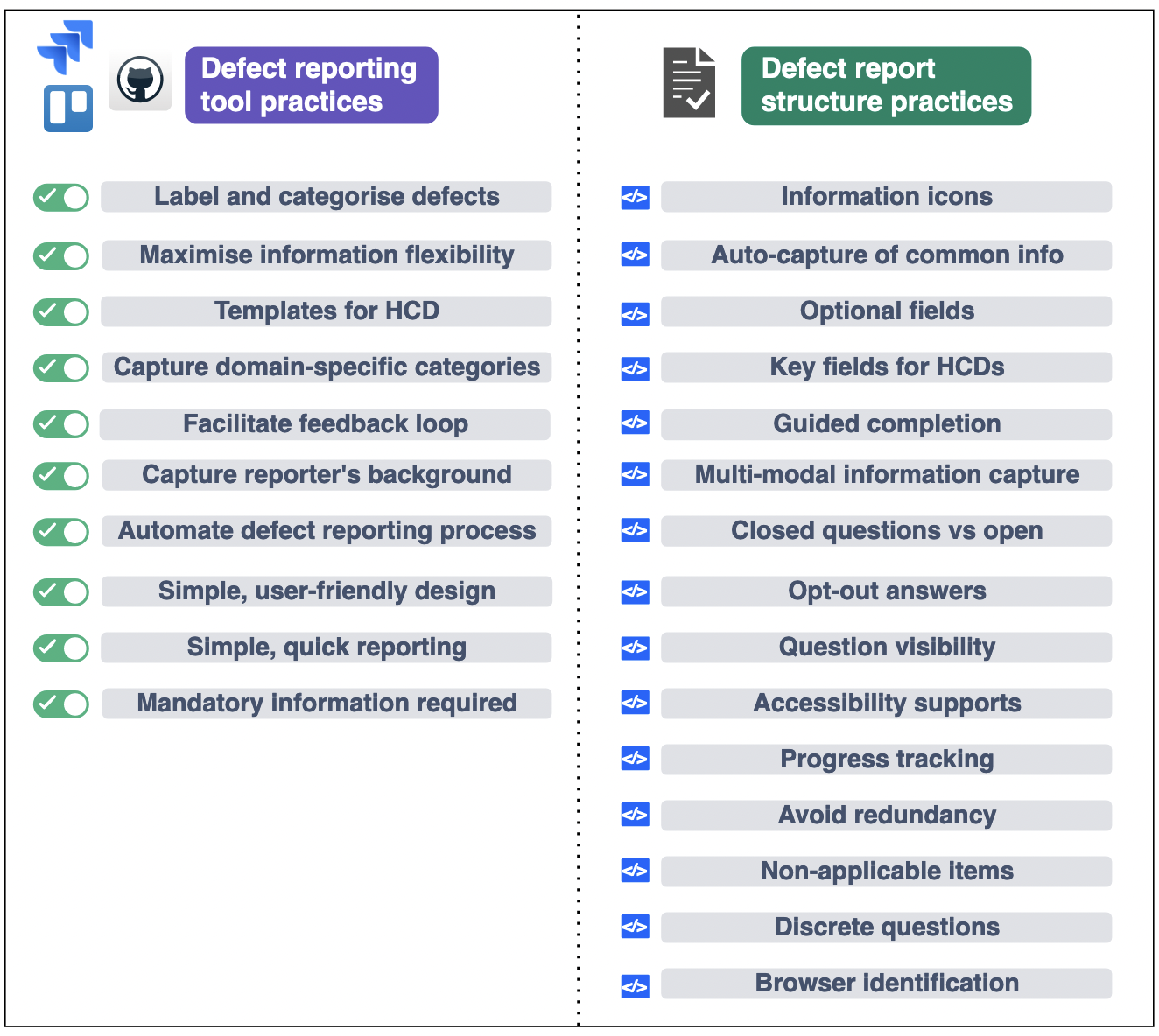}
\caption{Practices for Defect reporting tools and Defect report structure}
\label{fig:defectpractices}       
\end{figure*}

\subsubsection{Defect reporting tool practices.} 

Based on our discussions with practitioners, we created a list of practices a defect reporting tool should accommodate to capture HCDs.

\begin{itemize}
\item \textbf{Label and categorise defects:} The defect reporting tools should have options to add labels or categorise errors based on the domain or category of defects.
IP-5 claimed, \textit{``If you add labels, it narrows down the scope for a developer to actually drill down and lead to that person's defect."}

\item \textbf{Maximise information flexibility:} The tools should aim to maximise the amount of information captured, which can help developers in addressing the defects. However, in an attempt to do that, users should still have the flexibility to provide only the necessary information.
IP-1 suggested, \textit{``Collecting necessary information as much as possible from the user beforehand can help the developer solve the issue such as users providing videos and audios as an explanation of their defects would be much faster than writing the defects."}

\item \textbf{Templates for HCDs:} The tools may include a templating feature for HCDs, allowing for the pre-fill of some information in advance.
IP-2 highlighted, \textit{``There are different types of human-centric errors. Therefore, creating a template to add that information in an appropriate format would be ideal i.e. a universal guide for such defects."}

\item \textbf{Capture domain-specific categories:} The tool should capture domain-specific category information for the defect.
IP-3 claimed, \textit{``In the financial domain if the user experiences transaction troubles, they can provide such details to solve the problem. However, in healthcare, there might be patient information required rather than transaction data. Therefore, tools should capture such differentiating information."}
\item \textbf{Facilitate feedback loop:} The tool should provide a feedback loop for customers and developers to discuss their problems and help them provide the necessary information.
IP-1 suggested, \textit{``Not every user will provide detailed technical details. Therefore, a feedback loop where a user can provide more info after the defect is raised will be ideal for getting more details regarding a defect from both the end-users and the developer's side."}
\item \textbf{Capture reporter's background:} The tools should capture the reporter's background for a better understanding of users and the impact of the issue.
IP-6 stated, \textit{``A form in the application asking users to provide their information. This form should include details of why the data is collected because people won't fill in the details due to privacy concerns."}
\item \textbf{Automate defect reporting process:} Automation in these defect-reporting tools can help speed up the process of reporting issues.
IP-9 claimed, \textit{``Questions such as OS, version, hardware, etc. which can be avoided to collect information from the user should be automated to speed up the process."}
\item \textbf{User-friendly design:} The design should be simple and user-friendly for both technical and non-technical audiences.
IP-4 suggested, \textit{``Asking simple questions, less required questions, and keeping the overall design simple for both savvy and non-savvy users should be the aim of every tool."}
\item \textbf{Simple, quick reporting:} Reporting an issue should not be time-consuming and a simple structure can contribute to it.
IP-2 highlighted, \textit{``If I'm an end user, and I'm using the application and I find an issue, I will report it. If I find the second issue next year which is not a similar one but a different one, I shouldn't be filling in all the information regarding the profile and should be pre-filled, otherwise it will be very irritating. In short, the faster we report the better."}
\item \textbf{Mandatory information required:} There should be some mandatory information, otherwise the defect reports might be empty for most cases.
IP-7 suggested, \textit{``I think making the fields mandatory helps in better analysis. Consider a user trying to report a bug, making a few fields mandatory, has kind of meant that users think about, what sort of information they need to provide so that the report looks a bit more descriptive and easier to understand for a developer to analyse and review the problem."}
\end{itemize}

\subsubsection{Defect reporting structure.}

Based on the discussions with practitioners we created a list of practices a defect reporting template should accommodate to capture HCDs.

    \textbf{Information icons:} The defect report can have an information icon in front of the fields which the development team consider important such as product/component, Operating System (OS), Version, and Hardware.
    IP-11 highlighted, \textit{``You can always provide users with brief information and a concise description of how to obtain that information. For instance, when entering card details, users are routinely asked for the last three digits of their card. A pop-up or a help icon typically accompanies this request, explaining how to locate this information. Similarly, a help icon could describe the steps to determine the hardware version being used."}
   
    \textbf{Auto-capture of common info: } There should be automation for the fields such as OS, Hardware, and Version which can be captured on the backend without the user wasting time to fill these fields.
    IP-9 claimed, \textit{``It is challenging for users to determine the version they are using and where to find this information. I think it would be beneficial to include a link or instructions to assist users in locating this information. Ideally, this process could be automated to minimise user inquiries. If developers could automatically obtain this information, many questions could be avoided."}
    
    \textbf{Optional fields: } The defect reports should not make every field mandatory to fill by the user because it can lead to users not raising issues or adding random information in the fields. 
    IP-3 suggested, \textit{``I would suggest removing a few of these mandatory fields, such as the expected behaviour or observed behaviour fields. Often, people will simply enter "not available" or input something nonsensical. Therefore, making all of these fields mandatory does not seem reasonable to me."}
    
    \textbf{Key fields for HCDs: } Bug description, steps to reproduce, attachments, and impact are some of the important fields to capture from the HCD reporting perspective.
    IP-3 highlighted, \textit{``I would keep the description and attachments as a mandatory field because it provides essential details about the issue itself."}

    \textbf{Guided completion:} The defect reporting tool should provide guided support for non-technical users to complete HCD reports.
    For example, IP-5 stated, \textit{``If there were a form that specifically prompted users for information, I would include fields such as ``When did it occur?" and ``Provide a screenshot." It effectively captures all necessary details by asking, ``What do you see on the screen that is causing you problems?" and ``When did you see it?" This guided input method is undoubtedly beneficial."}

    \textbf{Multi-modal information capture: } Logs, timestamps, videos, and audios are also useful from a technical standpoint to solve HCDs.
    IP-5 claimed, \textit{``I require a timestamp and a screenshot of the issue. At a minimum, I ask users to provide this information if they are experiencing problems. However, I do not believe it is primarily the user's responsibility to do this. It is our duty to guide them and specify the exact information we need."}

     \textbf{Closed questions vs open: } Providing users with more close-ended questions in the form of drop-downs, radio buttons or checkboxes can help in getting more information.
    IP-10 suggested, \textit{``Consider using radio buttons or similar elements that are easier to navigate and more visually accessible. This relates to user interface design. It would be beneficial to ask questions one by one rather than overwhelming users with multiple questions at once."}

     \textbf{Opt-out answers: } The report should not mandate the user to fill each field.
    IP-13 highlighted, \textit{``Users should have the option to select ``prefer not to say" and proceed to the next step. These details should not be compulsory; it is the user's choice whether to provide them or not."}

     \textbf{Question visibility: } In some tools, the questions are only visible as part of the inline textbox of the answer field, which can be inconvenient. Hence, the questions should be always visible.
    IP-4 claimed, \textit{``I believe it may be beneficial to present the questions upfront in a form, rather than as inline questions. By having separate answer boxes, users can easily refer back to the question while formulating their responses. This approach prevents the issue of questions being hidden while users are trying to answer them, which can lead to loss of context and potential forgetfulness among users who may not be fully attentive."}

     \textbf{Accessibility supports: } Implement accessibility-related adjustments in defect forms to accommodate diverse user needs while filling a defect report.
    IP-15 stated, \textit{``For some individuals, reading smaller fonts can be challenging, particularly when catering to non-technical users. Therefore, it is imperative to implement accessibility-related adjustments in defect forms to accommodate diverse user needs."}

     \textbf{Progress tracking: } There should be a progress bar visible for the user when filling in the defect report to provide an indication of the state of the defect report at every possible time to the user.
    IP-7 suggested, \textit{``It would be beneficial to include a progress bar at the top, indicating the number of completed steps. This feature would help users gauge the remaining tasks and prevent the feeling of endless progression when clicking `Next'."}

     \textbf{Avoid redundancy: } The duplicate/repeatable fields should be avoided in the report.
    IP-1 highlighted, \textit{``If a user is reporting multiple issues, they should not need to reiterate their information for each issue."}

     \textbf{Non-applicable items: } The designers should consider whether a given field is really mandatory or else should be made optional to answer.
    IP-2 stated, \textit{``There should be an option available for users to indicate ``not applicable" or ``not available." This feature would allow users to select this option if they lack certain information or are unsure how to obtain it, thereby ensuring that they are not restricted from submitting the bug report."}

     \textbf{Discrete questions: } The questions should be structured in a way which collects separate information and not be combined into one big question.
    IP-11 claimed, \textit{``Separating the question into distinct options could streamline this process and facilitate a more efficient analysis of user feedback."}

     \textbf{Browser identification: } If the issue is related to the browser being used, automatically capture information on the type, version, plug-ins etc of the browser being used by the user in the defect form to better identify  the issue.
    IP-13 highlighted, \textit{``I would also include the specific browser in which the issue is being encountered if it is browser-related."}

\begin{tcolorbox}[%
       arc=0mm,width=\columnwidth,
                  top=1mm,left=1mm,  right=1mm, bottom=1mm,
                  boxrule=1pt, breakable]
\textbf{RQ4 Key findings.} 
\begin{itemize}
\item Practitioners shed light on current practices for defect reporting tools and defect reporting structure for capturing HCDs efficiently.
\item Some practices include labels, categorisation of the errors, templates, feedback loop with customers, customer support, simple and user-friendly design, and capturing important information to solve HCDs.
\item Defect report structure needs improvement and can be improved by utilising automation, information tags, avoiding repeatable fields, capturing customer information, optional sections, and close-ended questions for better understandability.
\end{itemize}
\end{tcolorbox}
\vspace*{.5em}


\section{Discussion}
\label{sec:discussion}

\subsection{Classification of HCDs}
\label{subsec:discussionclassification}

In our previous study, practitioners highlighted that HCDs are frequently reported in the usability category followed by functional, compatibility, performance, and security defect categories \citep{chauhan2024software}. In order to collect evidence on the claims established by practitioners, we classified defects in GitHub. Additionally, we discussed an example from each of the five key categories with 15 practitioners to provide validation on the HCDs in the following categories. Based on the results, \textbf{we have identified that the functional defect category has more HCDs followed by usability, performance, security and compatibility.} Both functional and usability have more defects, as both aspects are core to the software applications and to interaction from an end users' standpoint~\citep{rodriguez2014usability}. 

\textbf{After analysing 1,100 issues, we identified only 16\% of HCDs.} This shows that tools such as GitHub do not provide a formal medium to end-users for reporting defects. Due to the lack of HCD awareness, the discussions between software practitioners in the SE industry around reporting and fixing HCDs are limited. \textbf{We identified very nuanced discussions and the defect categorisation we have created is based on those discussions.} There needs to be extensive data collection to generate an exhaustive defect categorisation. For instance, we have identified HCDs in functional and usability defects categories comprising around 90\%. Therefore, the rest of the defect categories should be explored further in the creation of the defect categorisation. As part of the future research directions, the defect categorisation can be evolved into more detailed factors while considering HCDs. \citep{khalajzadeh2022diverse,houriehhcdinapps} explored app reviews and issue comments to provide a distinction between end-users and developers. Our research study validates the findings of their research as there is less discussion by developers on human-centric issues. Additionally, our study focuses more on the traditional defect categories as compared to more specific human-centric issues discussed by \citep{khalajzadeh2022diverse,houriehhcdinapps}.

\textbf{We also identified that HCDs are different across projects, domains, and categories to defects.} For instance, a functional defect in the ``system-design-primer" on the link excluded for documentation could also relate to usability issues if the link is broken. IT-Manufacturing, IT-Web applications and IT-Healthcare have around 66.5\% of HCDs, particularly the functional and usability side. IT-Spatial and IT-Gaming have the least HCDs captured i.e., 19.3\%. We only captured 100-200 issues per domain. This is fairly less and hence, the domain-level and defect categories attributes captured require more extensive data collection and discussion on how the software practitioners discuss HCDs in these areas. For instance, security-related HCDs are 2.3\% and understanding whether this domain has HCDs or not can be better captured by utilising security-related projects. This encourages new research areas to capture and explore HCDs fixing, frequently reported HCDs, domain-specific HCDs, and HCDs in different defect categories.

\citep{houriehhcdinapps} identified that human-centric issues can be both technical and non-technical. Their category division identified the perception of technical issues and human-centric issues such as compatibility usually represented as technical but it could be human-centric due to the social-economic status of end-users not able to afford the latest technology. \textbf{We identified that the HCDs are more technical in open-source repositories.
} The issues raised discuss technical details and very limited non-technical details. Hence, the definition of HCDs is identified as nuanced defects that are missing edge conditions. For instance, a user requiring documentation update in the application is dealt with more technical details and can only be interpreted by the developers of the application, \textit{``I wanted to use the library and discovered the installation section of the documentation. The only documented way is conda, even though the library is published and up-to-date on Pipy. It didn't take me long to figure it out but that was still a missing information. I would like the documentation to be up-to-date."} - [osmnx].

\textbf{While classifying the HCDs, we identified the defect reporting tools, especially open-source tools such as GitHub do not provide a proper defect reporting structure to report the defects.} Additionally, the templates do not include formal defect reporting fields such as attachments, descriptions, steps to reproduce, impact, severity, and so on. Some domains have simple templates such as IT-Healthcare, while some have complex templates such as IT-Web applications. The reporting process is very convoluted; hence most end-users don't use them and prefer to leave reviews. \citep{yusop2016reporting} propose utilising a wizard-based approach as the preferred method for reporting usability problems. Furthermore, an improved severity, prioritisation, and customised reporting format are needed to accommodate various sorts of usability issues. To handle HCDs, these defect reporting tools require better classification, reporting structure, and overall revamp of the reporting process. Hence, more related research is required in this domain.

\subsection{Defect reporting process and reporter’s background}

\textbf{Defect reporting tools such as Jira and GitHub have a generalised structure to capture defects.} In order to capture HCDs in different domains and defect categories, these tools need to be revamped to account for these changing domain and defect categories paradigms. HCDs in the domains and defect categories are subjective; hence, one size fits all would not be able to capture such defects. Therefore, the inclusion of domain-specific categories, labels, tags, and the specific impact on the end-user should be the goal. We need to account for changes in severity and priorities based on the user impact. For instance, \citep{yusop2016reporting} proposed that the inclusion of an improved severity, prioritisation, and customised reporting format is needed to accommodate various sorts of usability issues. A similar structure with an addition to the defect categorisation could effectively capture the HCDs.

\textbf{There should be proper documentation of the domain defect reporting structures.} Any minor or major change across the domains and defect categories should be documented in relation to HCDs. This can help in learning new HCDs as well as training software development teams to manage them. A potential research area is the creation of such a taxonomy, we have created a preliminary defect categorisation that can be broken down into granular levels of domains and defect categories. In their studies, \citep{yusop2016reporting,yusop2020revised} propose that usability defects lack a uniform taxonomy. They suggest the development of a taxonomy that can accurately describe and characterise the aspects of usability problems. This taxonomy would enhance the process of reporting, prioritising, comprehending, and resolving these issues.

Defect reporting tools will be utilised by the end-user experiencing a genuine problem. A problem that is subjective to the user could be a potential HCD. If the development teams prioritise the defects based on the number of users experiencing such problems, then they can lose the customer support in the process. If a single user encounters such an issue, it may not be surfaced among the vast number of technical defects. This could lead to a loss of customer business to potential competitors. If this happens more often, the business will identify the impact. Hence, empathy regarding the end users can help businesses solve their issues and eventually grow their target audience. \textbf{Integrating the reporter's background in defect reporting tools can help in creating better defect reports for developers to understand the issue from the end-user perspective.} Developers can empathise with end-users' struggles based on their background and create healthy discussions in teams to solve the issues. These teams can identify their target audience and learn the problems faced by the end-users. It could be an extension of end-user testing practices. However, most end-users could be hesitant to provide their private information in the report. An incentive can be provided to the users to mitigate privacy issues from these users' background information. Providing concise information about why the information is getting collected, how the information will be used, and the advantage of providing such information can help in concrete defect reports showcasing impact to the end-user. However, such information collection should be optional and handled sensitively. This can help in creating a long-term meaningful relationship with the end-user~\citep{grundy2020humanise}.

\textbf{Automation, simplicity, timeliness, and intuitiveness are crucial qualities of an HCD tool.} Human-centric domain experts can be utilised in creating templates, improvements in tools, end-user defect reporting, and automation scope in the tools. Collecting necessary information without wasting much time should be the primary focus. The practices provided by the practitioners could commence managing HCDs efficiently. The ideal structure of the defect reports and tools from the practices could eventually capture HCDs in an intended manner. Existing defect reporting tools lack such practices. Therefore, a future research direction could be to explore such avenues and enhance the management of HCDs. Generative AI models are becoming prevalent and usage of such models in defect reporting especially human-centric issues defect reporting could be helpful. They can help automate, tag HCDs to appropriate categories, speed up the reporting and provide customer support to the struggling end-users. The defect reporting tools with the combination of the practices and GenAI models could be the ideal approach for capturing HCDs. \citep{plein2023can} provides the usage of ChatGPT\footnote{https://chat.openai.com} in understanding and reproducing the bug reports. Their research showcases that ChatGPT automatically addressed half of the reported bug reports. This inclusion can extremely help in reporting and fixing of HCDs. Hence, more research needs to be conducted in this area.

\subsection{Recommendations for SE practitioners}

We identified a set of recommendations that practitioners can consider to include in their existing defect-reporting tools to improve the reporting of HCDs. \textbf{Utilising the reporter's persona} or profile in solving defects could help development teams understand issues from the end-users' perspective. This could be a practice conducted as part of end-user testing practices to solve such human-centric problems.
HCDs could be \textbf{better identified in defect reports} inside tools. Defect reporting tool usage practices could be improved by the software development teams to better identify and capture HCDs. This could include adding labels, categorisation HCD-related errors, using different categories of HCD defect report templates, reporting HCD resolution works in a feedback loop with customers, and creating simple and user-friendly report template designs to manage HCDs.
HCDs can be reported better by employing an \textbf{enhanced defect report strategy and structure} by development teams. This can include utilising more HCD defect handling automation, HCD  information tags, avoiding repeatable fields, capturing customer and defect reporter information, use of optional sections potentially irrelevant for some HCDs, and close-ended questions for better understandability.
Finally, different \textbf{HCD experts from different domains} can create HCD-related defect categorisation lists and practices to better manage HCDs. For example, HCDs specifically relevant to eHealth apps could be identified with suitable tags, fields, and automation to capture richer eHealth HCD information.

\subsection{Future research}

The classification we conducted in the study includes a very limited amount of data. Researchers can capture more data by analysing existing open-source or closed-source defects to create a more exhaustive taxonomy of HCDs in defect reporting categories. Additionally, they can explore other functional and non-functional defects in detail to verify the state of HCDs in open-source repositories. Utilising such data, detailed ML or GenAI models can be created to effectively assist the classification of HCDs in the SE domain.

Better defect reporting, triage, and analysis automation are required to assist HCD management in reporting and fixing issues more effectively. Researchers can utilise our HCD classification to explore additional categorisation and handling for HCDs. The defect reporting structure and tools could be improved by applying the practices discussed in this research study to enhance the development teams' ability to capture and fix HCDs. Additionally, the usage of GenAI in these defect reporting tools could be explored to efficiently capture HCDs as well as issue summarisation for assisting the fixing of HCDs.

Additionally, conducting a focused study with the specific organisation can help determine the problems SE practitioners face in capturing and fixing HCDs. Researchers can create specific test scenarios to understand the problems and create a detailed taxonomy and practices for HCDs.

\section{Threats to validity}
\label{sec:threats}

\sectopic{Internal Validity.} The selection of open-source projects and corresponding issues could introduce internal bias. The dataset of 11 projects and 100 issues per project might be too small to capture the full range of HCDs. Although we randomized the collection of these 1,100 issues, manually analyzing all issues from these projects was not feasible. Attempts to automate the classification process were unsuccessful, which could have led to missed projects and issues, introducing potential threats to internal validity.

The defect categorisation we developed for HCDs in defect categories may be subjective and susceptible to errors. The first three authors of the study helped in the qualitative analysis of the defect categorisation. Any disagreements between the authors were dealt with in an open discussion, and they came to an agreement based on the discussions. Any issues that we were not confident were placed under non-HCD buckets. We could have missed some HCDs; however, we are confident that we captured the defect categorisation based on the collected data with minimal mistakes.

Another challenge to internal validity is whether participants understood the concept of HCDs in the same manner as we intended. The reliability of the results would be compromised if practitioners lacked a thorough understanding of the subject matter or employed alternative interpretations although possessing a functional understanding of defect management in the SE domain. In order to mitigate any potential confusion among participants, we provided explicit explanations and illustrative instances for each topic. We consider this threat to be negligible as we have not detected any instances of misunderstandings or contradictions in the responses provided by the practitioners. Additionally, distributing gift cards to the interview participants could provide another potential risk. Nevertheless, we approved the compensation for the participants after carefully evaluating their responses to ensure that they answered every question and met the criteria for our target audience.

\sectopic{Construct Validity}
The primary threat to construct validity concerns the potential for practitioners to misinterpret the interview questions. To ensure clarity, we designed the questions carefully and conducted a pilot study with research professionals experienced in software development and testing. We also provided practitioners with adequate examples to prevent bias, allowing them to respond based on their judgment.

\sectopic{Conclusion Validity.}
Additional data availability may lead to other inferences being drawn from the information obtained in this investigation. We depend on the assessments of practitioners regarding the defect reporting process and practices, which may not accurately reflect the actual situation.

\sectopic{External Validity.}
The results from the classification could be less generalisable due to the small dataset of 11 projects and 1,100 issues. Additionally, GitHub as a defect reporting tool is much more prevalent to technical audiences such as developers and reporters rather than end-users. Hence, HCDs are much more technical based on those discussions. A more comprehensive set of HCD defect categorisation and taxonomy can be created by analysing other open-source software repositories such as Bugzilla and Bitbucket, application reviews from Play and App stores, and software artefacts (e.g., commits, requirement specifications) of proprietary and open-source projects. In addition, the defect reporting tool practices produced based on the practitioner's opinion may be less comprehensive due to the limited number of personnel, such as 15. Surveys or focus studies in the SE domain can allow us to gather a broad range of perspectives and information.


\section{Conclusion and Future Work}
\label{sec:conclusion}

HCDs are nuanced and subjective defects that are often generated from users' perceptions or missed by development teams in either development or testing. In our previous study, we identified HCDs suffer from a lack of awareness across the SE industry and hence, the existing practices supporting defects could not account for HCDs in their software. Development teams lack comprehension of these defects, resulting in user queries being overlooked. Defect reporting tools lack the proper management of HCDs. Hence, this research explores the classification and management of HCDs in defect reporting tools. 

We manually classified 1,100 open-source defects from the GitHub defect reporting tool and identified 176 HCDs across 6 different domains, such as IT-Healthcare, IT-Web, IT-Spatial, IT-Manufacturing, IT-Finance, and IT-Gaming. We identified IT-Manufacturing had the highest number of HCDs. Additionally, the functional defect category had more HCDs, followed by usability, performance, security and compatibility. Based on the classification, we created a preliminary defect categorisation of HCDs in defect reporting categories i.e. in functional and non-functional defects.

We identified the reporting structure in GitHub does not provide ideal reporting of HCDs. In order to validate the results of our classification and improve the defect reporting process, we interviewed fifteen software practitioners, i.e., developers and reporters, to provide transparency on the management of HCDs in existing defect reporting tools. We comprehended the existing defect reporting tools do not provide a proper defect report structure to HCDs. These tools can not capture domain and defect category information relating to HCDs. Additionally, we identified the process of defect reporting requires priority of the defect based on the number of users impacted by the defect. In the case of HCDs, we could lose the defect because it is very subjective and does not necessarily fit one size at all.

Based on practitioners' comments, we determined that integrating the reporter's background in defect reports could be useful to help developers empathise with the end-users and understand the issue from their perspective, thereby increasing the understandability of the defect report. We, therefore, create a set of practices required by an ideal defect-reporting tool to manage HCDs. If the tool captures most of the parameters provided by the practitioners, then the development teams can solve HCDs more effectively.
 
As part of future work, we plan to explore ML models or Generative AI models to automate the classification of HCDs. Additionally, we would attempt to improve the defect categorisation of HCDs in defect reporting categories by capturing more diverse projects, repositories, application reviews, and software artefacts such as requirement specifications and commits. Additionally, improvements in the defect reporting tools can be explored to enhance the process of capturing the reporting and fixing of HCDs. 

\vspace*{0.5em}
{\sectopic{Acknowledgment.} Chauhan and Grundy are supported by ARC Laureate Fellowship FL190100035.}\\

\sectopic{Declarations} 
{\sectopic{Conflict of Interest.} The authors declare that they have no conflict of interest.}
{\sectopic{Data Availability Statement.} The manuscript has associated data in Zenodo repository~\citep{anonymous_survey_and_interview}}.

\balance
\bibliographystyle{spbasic}

\bibliography{paper}

\end{document}